\documentclass[aps,prd,twocolumn,preprintnumbers,nofootinbib,amsmath,amssymb,superscriptaddress,10pt, floatfix]{revtex4-2}

\usepackage{graphicx}
\usepackage{placeins}

\usepackage{hyperref}
\usepackage[capitalise]{cleveref}
\crefname{section}{Sec.}{Sections} 

\usepackage[font=small, labelfont=bf, textfont={small,it}]{caption}
\usepackage{multirow}
\usepackage{makecell}
\usepackage{booktabs}

\usepackage{xcolor}
\usepackage{xspace}
\usepackage[normalem]{ulem}

\newcommand{\be}{\begin{equation}}
\newcommand{\ee}{\end{equation}}
\newcommand{\benn}{\begin{equation*}}
\newcommand{\eenn}{\end{equation*}}

\newcommand{\tzero}{$t_0$\xspace}
\newcommand{\tzeroproj}{$t_0^{(0)}$\xspace}
\newcommand{\tone}{$t_1$\xspace}
\newcommand{\wzero}{$w_0^2$\xspace}
\newcommand{\wzeroprime}{$w^{\prime 2}_0$\xspace}
\newcommand{\flow}{$\varphi(t)$\xspace}
\newcommand{\flowprime}{$\varphi^{\prime}(t)$\xspace}
\newcommand{\qzero}{$Q=0$\xspace}
\newcommand{\SU}[1]{$\mathrm{SU}(#1)$}
\newcommand{\SUN}{$\mathrm{SU}(N)$\xspace}
\newcommand{\TBC}{TBCs\xspace}
\newcommand{\PBC}{PBCs\xspace}
\newcommand{\OBC}{OBCs\xspace}
\newcommand{\ov}{\scriptscriptstyle{\mathrm{ov}}}
\newcommand{\clov}{\scriptscriptstyle{\mathrm{clov}}}
\newcommand{\Tr}{\mathrm{Tr}}
\newcommand{\TGF}{\scriptscriptstyle{\mathrm{TGF}}}
\newcommand{\leff}{l_s^{\scriptscriptstyle{(\rm eff)}}}

\newcommand{\Veffc}{{\cal V}_{\scriptscriptstyle{\rm eff}}}
\newcommand{\Vc}{{\cal V}}
\newcommand{\muhad}{\mu_{\scriptscriptstyle{\rm had}}}
\newcommand{\dd}{\mathrm{d}}
\newcommand{\eee}{\mathrm{e}}
\newcommand{\ii}{\mathrm{i}}
\newcommand{\minf}{m_{\scriptscriptstyle{\infty}}}

\begin{document}

\title{Scale setting of SU(\ensuremath{N}) Yang--Mills theory, topology and large-\ensuremath{N} volume independence}

\author{Claudio Bonanno}
\email{claudio.bonanno@csic.es}
\affiliation{Instituto de F\'isica Te\'orica UAM-CSIC, c/ Nicol\'as Cabrera 13-15, Universidad Aut\'onoma de Madrid, Cantoblanco, E-28049 Madrid, Spain}

\author{Jorge Luis Dasilva Gol\'an}
\email{jgolandas@bnl.gov}
\affiliation{Physics Deparment, Brookhaven National Laboratory, Upton, New York 11973, USA}

\author{Margarita Garc\'ia P\'erez}
\email{margarita.garcia@csic.es}
\affiliation{Instituto de F\'isica Te\'orica UAM-CSIC, c/ Nicol\'as Cabrera 13-15, Universidad Aut\'onoma de Madrid, Cantoblanco, E-28049 Madrid, Spain}

\author{Massimo D'Elia}
\email{massimo.delia@unipi.it}
\affiliation{Dipartimento di Fisica, Università di Pisa \& INFN, Sezione di Pisa, Largo Pontecorvo 3, I-56127 Pisa, Italy}

\author{Andrea Giorgieri}
\email{andrea.giorgieri@phd.unipi.it}
\affiliation{Dipartimento di Fisica, Università di Pisa \& INFN, Sezione di Pisa, Largo Pontecorvo 3, I-56127 Pisa, Italy}

\date{\today}

\begin{abstract}
We set the scale of SU($N$) Yang--Mills theories for $N=3,5,8$ and in the large-$N$ limit via gradient flow, as a first step towards the computation of the large-$N$ $\Lambda$-parameter using step scaling. We adopt twisted boundary conditions to achieve large-$N$ volume reduction and the Parallel Tempering on Boundary Conditions algorithm to tame topological freezing. This setup allows accurate determinations of the gradient-flow scales down to lattice spacings as fine as $\sim 0.025$ fm for all the explored values of $N$, a regime that has never been reached with ergodic algorithms. Moreover, we are able to precisely estimate the finite-size systematics related to topological freezing, and to show the suppression of finite-volume effects expected by virtue of large-$N$ twisted volume reduction.
\end{abstract}

\maketitle

\section{Introduction}

In the last decade, it has become increasingly common for lattice studies of QCD and QCD-like systems to reach percent or even subpercent levels of accuracy. Recent paramount examples are the lattice computation of the hadronic contribution to the muon anomaly (see the review~\cite{Aliberti:2025beg} and references therein) or the lattice computation of the strong coupling constant (see Ref.~\cite{Brida:2025gii} and references therein). As a higher precision is achieved, new possible sources of systematic effects must be investigated and controlled. In this regard, \emph{scale setting} plays a crucial role, as its uncertainties and systematics propagate to any observable which needs to be converted from lattice to physical units.

Gradient flow~\cite{Narayanan:2006rf,Luscher:2009eq,Luscher:2010iy,Lohmayer:2011si} scales have become a standard choice for scale setting, requiring only the numerical integration of the gradient flow equation and computation of the energy density on the lattice. These are considered to be reliable reference scales, as numerical integration can be performed with arbitrarily small systematic error, and the energy density can be determined with high statistical precision. However, it is widely acknowledged that the scale-setting procedure via gradient flow may be subject to a bias due to finite-volume effects and \emph{topological freezing}~\cite{Fritzsch:2013yxa,RamosMartinez:2023tvx}. Topological freezing~\cite{Alles:1996vn,DelDebbio:2004xh,Schaefer:2010hu} constitutes a computational challenge that pervades Monte Carlo simulations of all lattice field theories possessing nontrivial topological properties when employing conventional local algorithms in the vicinity of the continuum limit. As the lattice spacing $a$ decreases, the Markov chain tends to remain trapped in a fixed topological sector, thereby losing ergodicity. It is well-known that expectation values taken at fixed topology suffer from powerlike finite-volume effects~\cite{Brower:2003yx,Aoki:2007ka}, thus a fixed topology enhances finite-size corrections with respect to the usual exponentially suppressed ones. Moreover, after the gradient flow, the energy density used to extract gradient-flow scales becomes highly-correlated with the topological charge of the underlying gauge fields. Thus, the ergodicity problem due to the incorrect topological sampling could also introduce unwanted biases in the scales themselves~\cite{Fritzsch:2013yxa}.

The goal of this paper is to perform the scale setting of large-$N$ Yang--Mills theories via the gradient flow on lattices with lattice spacings as fine as $\sim 0.025$ fm. Our main motivation is related to the fact that this is a necessary first step towards the determination of the large-$N$ $\Lambda$-parameter via step-scaling~\cite{Luscher:1991wu} from a finite-volume renormalization scheme based on the use of twisted boundary conditions~\cite{tHooft:1979rtg}. This is an ongoing project we are carrying on~\cite{Bribian:2021cmg,Golan:2021vja,DasilvaGolan:2023yvg,Bonanno:2024nba}. In the regime we are interested in, systematic effects related to topological freezing are particularly relevant (see below), and require to be addressed with particular care. Pursuing our principal goal thus requires to quantitatively characterize the behavior of systematic effects on gradient-flow scales in the presence of a frozen topology. Such topic is of much broader interest, and goes beyond the study of scale setting in large-$N$ gauge theories. Hence, our investigation may also be a useful reference on the subject even for communities working on other lattice field theories.

\begin{figure}[!t]
\centering
\includegraphics[scale=0.32]{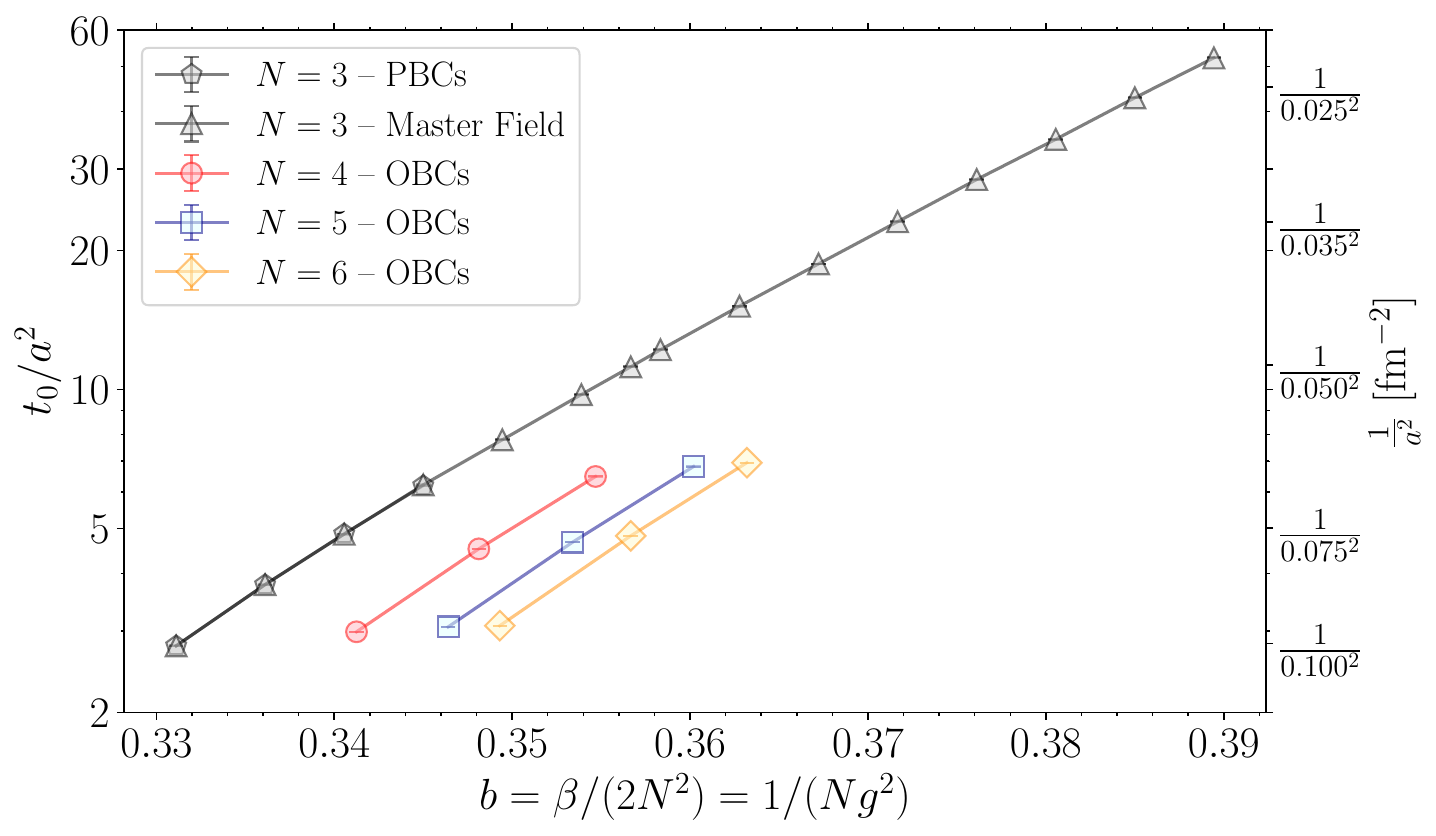}
\caption{Status of determinations of the popular gradient-flow scale $t_0$ in lattice units [see \cref{sec:gradflow_scales_def} for its definition] in large-$N$ Yang--Mills theories. Data for $N=3$ come from Ref.~\cite{Ce:2015qha} (\PBC) and Ref.~\cite{Giusti:2018cmp} (Master Field), while those for $N=4,5,6$ come from Ref.~\cite{Ce:2016awn} (\OBC). The right $y$-axis uses $\sqrt{8t_0}=0.475$ $\mathrm{fm}$ for every $N$ to convert the lattice spacing in physical units. The $x$-axis reports the inverse 't Hooft coupling $b=1/\lambda=\beta/(2N^2)$, with $\beta$ the standard inverse gauge coupling.}
\label{fig:t0_summary}
\end{figure}

\begin{figure}[!t]
\centering
\includegraphics[scale=0.32]{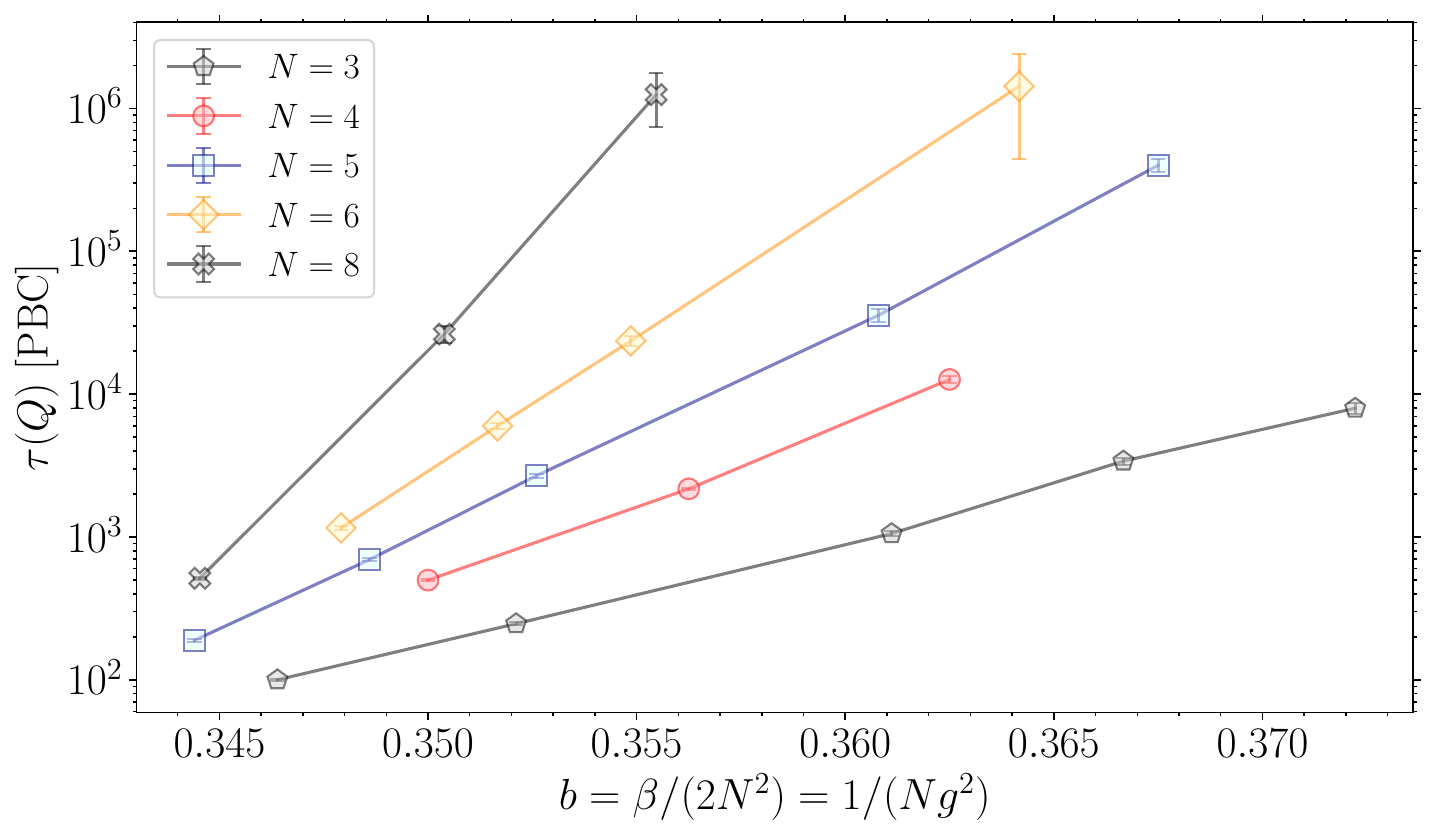}
\caption{Behavior of the integrated autocorrelation time of the lattice topological charge $\tau(Q)$ as a function of the inverse 't Hooft coupling $b=1/\lambda=\beta/(2N^2)$, with $\beta$ the standard inverse gauge coupling. These data, obtained from standard \PBC simulations, come from Ref.~\cite{Athenodorou:2021qvs}, and refer to the lattice clover topological charge computed after cooling.}
\label{fig:tau_PBC_vs_b}
\end{figure}

To better contextualize our study, it is useful to briefly review the existing determinations of gradient-flow scales in \SUN Yang--Mills theories. 
The case $N=3$ has been studied in most detail~\cite{Luscher:2010iy,Ce:2015qha,Francis:2015lha,Knechtli:2017xgy,Giusti:2018cmp}. However, only one paper~\cite{Giusti:2018cmp} determined gradient-flow scales below a lattice spacing $a \simeq 0.067$ fm with a definitive strategy to deal with topological freezing (namely, Master Field simulations). Concerning the case $N>3$, gradient-flow scale setting has been addressed in only one paper~\cite{Ce:2016awn} for $N=4,5,6$ and adopting open boundary conditions (\OBC)~\cite{Luscher:2011kk,Luscher:2012av} to alleviate topological freezing. There, only results down to $a \sim 0.064$ fm have been reported. We have collected all pure-gauge determinations of the lattice spacing in units of $t_0$ discussed so far in \cref{fig:t0_summary}, where $t_0$ is displayed as a function of the inverse bare 't Hooft coupling $b=(Ng^2)^{-1}=\beta/(2N^2)$, with $\beta$ the standard inverse gauge coupling. For illustrative purposes, the conversion of $a$ to physical units was performed assuming $\sqrt{8t_0}=0.475$ fm~\cite{Giusti:2018cmp} for every $N$. The present status is clearly insufficient for our goals --- requiring lattice spacings as fine as 0.025 fm --- and reflects the inherent difficulties in performing reliable simulations for $N>3$ even at coarse lattice spacings, due to the well-known worsening of topological freezing occurring when $N$ is increased. Such difficulties can be quantified by the scaling of the integrated autocorrelation time $\tau(Q)$, the number of lattice sweeps necessary to generate two decorrelated samples of the topological charge $Q$. In \cref{fig:tau_PBC_vs_b}, we show $\tau(Q)$ as a function of $b$ using the data of~\cite{Athenodorou:2021qvs}, obtained adopting Periodic Boundary Conditions (\PBC). Just as an example, consider the case $b\simeq 0.355$, corresponding to $a\simeq 0.052,\ 0.074,\ 0.087$ fm for $N=3,\ 5,\ 8,$ respectively. One finds $\tau(Q)\simeq 3\cdot10^2,\ 6\cdot10^3,\ 10^6$ in units of number of lattice sweeps. Despite the fact that \OBC alleviate topological freezing, the regime $N > 3$ and $a\lesssim 0.065$ fm still remains challenging and unexplored. For example, using the data of Ref.~\cite{Ce:2016awn} for the number of updating steps separating two subsequent decorrelated measures of the topological susceptibility, and expressing them in the same units used in~\cite{Athenodorou:2021qvs} as done in Ref.~\cite{Bonanno:2025eeb}, one can estimate $\tau(Q) \sim 1.5 \times 10^4$ lattice sweeps for $N = 6$ and $a \simeq 0.064$ fm (the finest lattice spacing explored in that study). On top of this, the larger temporal extent required to eliminate boundary effects adds an extra computational cost that should also be considered.

To pursue our goals, it is therefore clear that a novel numerical strategy must be devised. In this work, we adopt  the \emph{Parallel Tempering on Boundary Conditions} (PTBC) algorithm to effectively mitigate topology freezing, in combination with \emph{Twisted Boundary Conditions} (\TBC) to efficiently implement finite-size scaling. The PTBC algorithm is a well-established method of mitigating topological freezing in a variety of physical regimes (fine lattice spacings, large $N$, zero and finite temperature), both in $2d$ models~\cite{Hasenbusch:2017unr,Berni:2019bch,Bonanno:2022hmz}, and in $4d$ non-Abelian gauge theories with~\cite{Bonanno:2024zyn} and without~\cite{Bonanno:2020hht,Bonanno:2022yjr,Bonanno:2023hhp,Bonanno:2024ggk,Bonanno:2024nba,Bonanno:2025eeb} fermion content. Recently, the PTBC algorithm has been shown to outperform both \PBC and \OBC simulations for what concerns the autocorrelation time of $Q$ at fixed numerical effort, and has been employed to compute the topological susceptibility for lattice spacings much finer than those reached with these strategies~\cite{Bonanno:2025eeb}. The choice of \TBC over standard \PBC is instead driven by the concept of \emph{twisted volume reduction}~\cite{Gonzalez-Arroyo:1982hyq,Gonzalez-Arroyo:1982hwr,Gonzalez-Arroyo:2010omx} (Refs.~\cite{GarciaPerez:2014cmv,GarciaPerez:2020gnf} provide comprehensive reviews on this subject). It has been demonstrated that, under certain conditions that are satisfied thanks to \TBC, Yang–Mills theories enjoy a dynamical equivalence between color and space-time degrees of freedom in the large-$N$ limit~\cite{PhysRevLett.48.1063}. In the context of a finite $N$, and when considering specific choices of the twist, the finite-volume effects associated with an $L \times L$ lattice plane with \TBC should be equivalent to those with \PBC and $L \to NL$, up to $1/N^2$ corrections. This enables a reduction in the volume of the simulated lattices as $N$ increases, whilst maintaining control over finite-volume effects.

In the following, we present scale setting results for $N=3,5,8$ down to lattice spacings as fine as $a \sim 0.025$ fm, focusing on the values of $b$ that are necessary for our forthcoming calculation of the large-$N$ $\Lambda$-parameter via step-scaling. More precisely, we determine $t_0$ avoiding topological freezing thanks to PTBC for several lattice volumes and bare couplings in order to perform a detailed finite-size scaling study and quantitatively characterize finite-size corrections, both for ergodic and for fixed-topology simulations. This is a crucial point to attain our goals, as it allows the explicit subtraction of finite-size effects even in extreme cases where it is not feasible to simulate lattices large enough to avoid finite-volume effects. Our investigation allows also to check the convergence of ergodic and fixed-topology simulation data to the same value in the infinite-volume limit.

It should also be stressed that, while there are practically no previous results to compare with for $N=5,8$, our scale-setting results for $N=3$ have instead been computed with the goal of comparing with the Master Field simulations of~\cite{Giusti:2018cmp}. Since the Master Field approach has its own set of assumptions (e.g., possible difficulties in defining thermalization, possible issues with the diffusion of local topological fluctuations), checking the agreement with results obtained with an ergodic algorithm that explores different topological sectors is a significant accomplishment, that is first achieved in this study.

This paper is organized as follows. In \cref{sec:gradflow_scales_def} we define the several gradient-flow scales we will compute. In \cref{sec:setup} we describe our lattice discretization, Monte Carlo algorithm and numerical strategies. In \cref{sec:results} we present our numerical results for $N=3,5,8$ and in the large-$N$ limit. Finally, in \cref{sec:conclusions} we draw our conclusions and discuss future outlooks of this investigation.

\section{Scale setting with gradient flow}\label{sec:gradflow_scales_def}

The scale of \SUN Yang--Mills theories can be conveniently set using the gradient flow~\cite{Narayanan:2006rf,Luscher:2009eq,Luscher:2010iy,Lohmayer:2011si}, a smoothing procedure that evolves the gauge fields $A_\mu(x)$ in a time $t$ according to the flow equation
\be\label{eq:wilson_flow}
\partial_t B_\mu (x, t) = D_\nu F_{\nu \mu} (x, t), \quad B_\mu (x, t = 0) = A_\mu (x)\, ,
\ee
where $D_\mu$ and $ F_{\mu \nu}$ are the covariant derivative and the field strength tensor of the flowed fields $B_\mu(x, t)$. The gradient-flow scale \tzero is defined for \SU{3} as~\cite{Luscher:2010iy}:
\be\label{eq:t0_su3}
\left.\langle t^2E(t)\rangle\right|_{t\,=\,t_0} = \frac{3}{10} = 0.3\, ,
\ee
where $E(t)$ is the energy density of the flowed gauge fields,
\be\label{eq:clover_density}
E(t) = \frac{1}{2} \Tr \left [F_{\mu \nu} (x, t)F_{\mu \nu} (x, t)\right]\, .
\ee
In physical units, this corresponds to $\sqrt{8t_0}\simeq 0.475$ fm~\cite{Giusti:2018cmp}. To generalize this definition to \SUN, consider that, in the small-$t$ perturbative limit, the scaling with $N$ of $t^2E(t)$ gives~\cite{Luscher:2010iy}
\be
\langle t^2E(t)\rangle\underset{N\to\infty}{\sim}\frac{N^2-1}{N}\, .
\ee
Thus, a definition of \tzero that remains finite in the large-$N$ limit and coincides with \cref{eq:t0_su3} for $N=3$ is given by~\cite{Ce:2016awn}
\be\label{eq:t0_sun}
\varphi(t_0) = \frac{9}{80} = 0.1125 \, ,
\ee
where
\be\label{eq:phi}
\varphi(t) \equiv \frac{N}{N^2-1}\langle t^2E(t)\rangle\, .
\ee
Analogously, one defines \wzero~\cite{BMW:2012hcm} in the large-$N$ limit through the logarithmic derivative of \flow,
\be\label{eq:w0_sun}
\left.t\frac{\dd\varphi(t)}{\dd t}\right|_{t\,=\,w_0^2} = \frac{9}{80} = 0.1125\, .
\ee
In addition to these two, we will also consider the scale \tone, defined as cutting \flow at half the threshold of \tzero~\cite{Butti:2022sgy}:
\be\label{eq:t1_sun}
\varphi(t_1) = \frac{9}{160} = 0.05625 \, .
\ee
The preference for one or other of these scales is dictated by the need to minimize finite-volume effects and/or lattice artifacts on the flow. Given that the gradient flow smears the gauge fields on a ball of radius $\sqrt{8t}$, a condition to minimize both effects at a given flow time $t$ would be  $a\ll \sqrt{8t} \ll l$, with $a$ the lattice spacing and $l\equiv L a$ the physical size of the lattice. 

The normalization of \flow chosen in \cref{eq:phi} cancels the leading and subleading scaling with $N$ predicted by perturbation theory. Alternative definitions, canceling only the leading $N$-dependence of the flow, can also be found in the literature~\cite{GarciaPerez:2014azn,Perez:2020vbn}:
\be\label{eq:t0_prime_sun}
\varphi^{\prime}(t^{\prime}_0) = \frac{1}{10} = 0.1 \, ,
\ee
\be\label{eq:t1_prime_sun}
\varphi^{\prime}(t_1^{\prime}) = \frac{1}{20} = 0.05 \, ,
\ee
and
\be\label{eq:w0_prime_sun}
\left.t\frac{\dd\varphi^{\prime}(t)}{\dd t}\right|_{t\,=\,w_0^{\prime 2}} = \frac{1}{10} = 0.1\, .
\ee
where
\be\label{eq:phi_prime}
\varphi^{\prime}(t) \equiv \frac{1}{N}\langle t^2E(t)\rangle\, .
\ee
These definitions, adopted in twisted-reduced models where $N\sim\mathcal{O}(10^2-10^3)$ is so large that $N^2-1\simeq N^2$, define scales with a finite large-$N$ limit too, but only coincide with the previous ones for $N=3$.

Finally, to study the effect of topology on the scale setting, we will also compute all the scales obtained when the flow is projected into the sector with topological charge $Q=0$. This is achieved replacing \flow with
\be\label{eq:phi_proj}
\varphi_{0}(t) \equiv \frac{N}{N^2-1}\frac{\langle t^2E(t)\delta_{Q,0}\rangle}{\langle \delta_{Q,0} \rangle}\, ,
\ee
with $\delta_{Q,n} =1$ when $Q=n$ and $0$ otherwise, and the same for \flowprime. As explained in the Introduction, a fixed topological sector should not affect the gradient-flow scales defined in the thermodynamic limit. However, taking into account both the projected and non-projected definitions enables us to identify the finite-volume effects introduced by a frozen or poorly sampled topology. 

\section{Numerical setup}\label{sec:setup}

In this section, we describe our lattice setup and the algorithm used for the simulations. We also present a general discussion on the expected lattice artifacts and finite volume effects on the flow.

\subsection{Lattice action and observables}

We discretize the pure-gauge \SUN theories for $N=3,5,8$ using the Wilson plaquette action on lattices with a long size $l\equiv La$ along the two directions $\mu=0,3$ and a short size $l_s \equiv L_s a < l$ along $\mu=1,2$. We impose \PBC along the long directions and \TBC~\cite{tHooft:1979rtg,Gonzalez-Arroyo:1982hyq} in the plane with two short directions. The lattice action is given by:
\be\label{eq:lattice_action_TBC}
S_{\scriptscriptstyle{\rm W}}[U] = -N b\sum_{x,\mu \, \neq \, \nu} Z_{\mu\nu}^*(x)\,  \Tr \left[ P_{\mu\nu}(x)\right],
\ee
where $b = 1/\lambda$ is the inverse 't Hooft bare coupling ($\lambda=Ng^2=2N^2/\beta$, with $\beta$ the usual lattice coupling) and $P_{\mu\nu}(x)$ is the plaquette,
\be
P_{\mu\nu}(x) = U_\mu(x) U_\nu(x+a\hat{\mu}) U_\mu^{\dag}(x+a\hat{\nu}) U^{\dag}_\nu(x) \, .
\ee
The factor $Z_{\mu\nu}(x)$ implements \TBC. For $\mu < \nu$,
\be\label{eq:twist_factor}
\begin{split}
Z_{\mu\nu}(x) = Z_{\nu\mu}^*(x) =
\begin{cases}
\eee^{\ii \frac{2 \pi k}{N}}, & \makecell{(\mu,\nu)=(1,2)\\ x_{\mu}=x_\nu=0} \,, \\
\\[-1em]
1, & \text{otherwise,}
\end{cases}
\end{split}
\ee
where $k$ is an integer coprime with $N$. We take $k=1,2,3$ for $N=3$, 5 and 8, respectively, for reasons that will become clear below.

This setup is analogous to the one of Refs.~\cite{Bribian:2021cmg,Bonanno:2024nba}, which was used to determine the renormalized strong coupling in the Twisted Gradient Flow scheme. In that instance, the aspect-ratio $l / l_s = N $ and $k$ were predetermined as part of the scheme definition. However, the gradient-flow scales that are of interest in this study are defined in the thermodynamic limit. Consequently, they should not be dependent on the lattice geometry or the boundary conditions. Nevertheless, their choice can be optimized to reduce finite-volume effects. In this work, the same choice of twist as in~\cite{Bribian:2021cmg,Bonanno:2024nba} is maintained, selecting $k$ and $N$ such that they are separated by two steps in the Fibonacci sequence. In the Twisted Gradient Flow scheme, this choice allows to circumvent the emergence of tachyonic instabilities in the large-$N$ limit that invalidates the property of reduction~\cite{Chamizo:2016msz}. 

Furthermore, in contrast to Refs.~\cite{Bribian:2021cmg,Bonanno:2024nba}, the aspect ratio $l / l_s$ is not fixed to the number of colors. Instead, a series of simulations are conducted, encompassing various combinations of $l$ and $l_s$. The objective is to quantify and control finite-volume effects, but also to analyze the reduction of finite-size effects in the short and twisted directions, expected from \TBC over \PBC. It is anticipated that finite size effects in $l_s$ will fully disappear in the limit of large $N$.

As dimensionless energy density on the lattice, we use the clover definition
\be \label{eq:Eclover}
a^4E(t) = E_{\clov}(t) = -\frac{1}{2} \Tr\left[C_{\mu\nu}(x,t)C_{\mu\nu}(x,t)\right] \, ,
\ee
where $C_{\mu\nu}(x,t)$ is the clover operator on the $(\mu,\nu)$ plane in the site $x$ evaluated after the gauge links have been evolved for a flow time $t$. It is defined as the anti-Hermitian and traceless part of:
\be
\begin{aligned}
\frac{1}{4}\, \bigg[& Z^*_{\mu\nu}(x) P_{\mu\nu}(x,t) + Z^*_{\mu\nu}(x-a\hat{\nu}) P_{-\nu\mu}(x,t)  \\   
&+ Z^*_{\mu\nu}(x-a\hat{\mu}) P_{\nu-\mu}(x,t)  \\   
&+ Z^*_{\mu\nu}(x-a\hat{\mu}-a\hat{\nu}) P_{-\mu-\nu}(x,t) \bigg],
\end{aligned}
\ee
where $U_{-\mu}(x, t) = U_{\mu}^{\dagger}(x-a\hat{\mu},t)$. 

The flow equation in \cref{eq:wilson_flow} is discretized and integrated with the adaptive third-order Runge--Kutta method described in Ref.~\cite{Fritzsch:2013je}. This procedure allows the determination of both  \flow and \flowprime for discrete times $T=t/a^2$. In order to ascertain a scale, a spline interpolation of \flow and \flowprime (or its derivatives for \wzero and \wzeroprime) and of its statistical uncertainty is performed, after which the time at which it intersects the threshold defining the scale is extracted.

\begin{figure}[!t]
\centering
\includegraphics[width=\columnwidth]{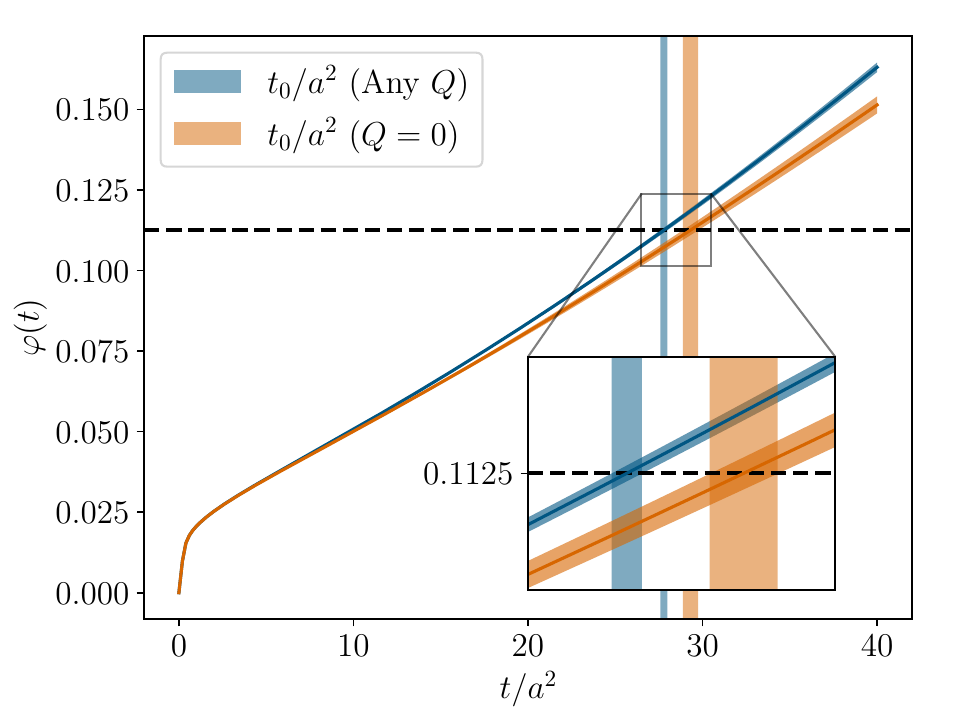}
\caption{Example of the determination of \tzero on an \SU{3} lattice with $L=38$, $L_s=24$, $b=0.37583$ and \TBC, with and without projection to $Q=0$.}
\label{fig:t0+_example}
\end{figure}

In order to study the effect of topology on the scale setting, one further ingredient is required; namely, a definition of the topological charge on the lattice. Given the clover discretization
\be
Q_{\clov}(t) = \sum_{x,\,\mu\nu\rho\sigma}\frac{\varepsilon_{\mu\nu\rho\sigma}}{32\pi^2}\Tr\left[C_{\mu\nu}(x,t)C_{\rho\sigma}(x,t)\right] \, ,
\ee
we assign to each configuration an integer topological charge
\be
Q = \mathrm{round}\left[Q_{\clov}(t_0)\right]\, .
\ee
We verified that $Q_{\clov}(t)$ reaches a plateau in $t$ before \tzero and is close to an integer number, thus it can be safely rounded to the closest integer. This allows to define the projection into the \qzero topological sector in \cref{eq:phi_proj}.
As an example, in \cref{fig:t0+_example} we show a lattice determination of \tzero, with and without topological projection.

\subsection{Monte Carlo algorithm}

The determination of the gradient-flow scales without resorting to the \qzero projection requires a correct sampling of the topological charge. As outlined in the Introduction, this is a complex task due to the occurrence of topological freezing at the fine lattice spacings that will be simulated. Topological freezing, moreover, further worsens when increasing $N$ at fixed lattice spacing. Open boundary conditions~\cite{Luscher:2011kk} or master field simulations~\cite{Luscher:2017cjh} have been employed to circumvent this issue in the case of \SUN pure gauge theories~\cite{Ce:2016awn,Giusti:2018cmp}. In this paper, the PTBC algorithm is adopted, following the implementation of Ref.~\cite{Bonanno:2024nba} with \TBC in the short directions. Originally proposed for two-dimensional $\mathrm{CP}^{N-1}$ models~\cite{Hasenbusch:2017unr}, the PTBC algorithm has been widely employed in the last few years to improve the state of the art of the lattice determination of several quantities. Various numerical studies have demonstrated that this approach leads to a substantial reduction in the autocorrelation time of the topological charge when compared with standard algorithms~\cite{Berni:2019bch,Bonanno:2022hmz,Bonanno:2024zyn,Bonanno:2020hht,Bonanno:2022yjr,Bonanno:2023hhp,Bonanno:2024ggk,Bonanno:2024nba,Bonanno:2025eeb}.

The main idea of the PTBC algorithm is to consider $N_r$ replicas $r=0,1,\dots,N_r-1$ of the lattice, each one differing for the boundary conditions imposed on a small subregion called the \emph{defect} $D$. We choose $D$ to be an $L_d^3$ spatial cube, placed on the time boundary $x_0 = L-1$. Links that cross $D$ orthogonally (i.e., temporal links) are multiplied by a real factor $c(r)$. For the physical replica (i.e., the one on which observables are computed) $c(0)=1$, so the defect has no effect and links enjoy PBCs. The other replicas interpolate between periodic and open boundary conditions on the defect: $c(N_r-1)=0$ for the last replica and $0 < c(r) < 1$ for those in-between. With \TBC, the defect is implemented by taking as the action of the replica $r$
\be
S_{\scriptscriptstyle{ \rm W}}^{\left(c(r)\right)}[U_r] = -b N\sum_{x,\mu \,\neq\, \nu} K_{\mu\nu}^{\left(c(r)\right)}(x) Z_{\mu\nu}^*(x)\Tr \left[ P^{(r)}_{\mu\nu}(x)\right] \, ,
\ee
where $U_r$ denotes the gauge links of the replica $r$. The factor $K_{\mu\nu}^{\left(c(r)\right)}(x)$ changes the boundary conditions on the defect, similarly to how the twist factor $Z_{\mu\nu}(x)$ in \cref{eq:twist_factor} implements \TBC:
\be
\begin{split}
K_{\mu\nu}^{(c(r))}(x) = &\enspace K_{\mu}^{(c(r))}(x) \cdot K_{\nu}^{(c(r))}(x+a\hat{\mu})\, \cdot \\
& \cdot K_\mu^{(c(r))}(x+a\hat{\nu}) \cdot K_\nu^{(c(r))}(x) \, ,
\end{split}
\ee
\be
K_{\mu}^{\left(c(r)\right)}(x) = 
\begin{cases}
	c(r),  & \mu=0\,,\,\, x \in D\, ,\\
	1 ,    & \text{otherwise.}
\end{cases}
\ee

For what concerns the Monte Carlo sampling, each replica is updated simultaneously and independently performing 1 lattice sweep of the standard local heat-bath algorithm~\cite{Creutz:1980zw,Kennedy:1985nu}, followed by $n_{\ov}=12$ lattice sweeps of the standard local over-relaxation algorithm~\cite{Creutz:1987xi}. Then, swaps among two adjacent replicas $(r,s=r+1)$ are proposed and accepted via a Metropolis step with probability
\be
p(r,s) = \min\left\{1, \eee^{-\Delta S^{(r,s)}_{\rm swap}}\right\} \, ,
\ee
\be
\begin{split}
\Delta S^{(r,s)}_{\scriptscriptstyle{\rm swap}} = & \,\,S_{\scriptscriptstyle{\rm W}}^{\left(c(r)\right)}[U_s] + S_{\scriptscriptstyle{\rm W}}^{\left(c(s)\right)}[U_r] \\
& -S_{\scriptscriptstyle{\rm W}}^{\left(c(r)\right)}[U_r] - S_{\scriptscriptstyle{\rm W}}^{\left(c(s)\right)}[U_s] \, .
\end{split}
\ee
The values $c(r)$ of intermediate replicas are tuned with short test simulations in order to achieve a mean acceptance $p$ of swaps around $20 \%$ for each pair of replicas. Thus, a given field configuration performs a sort of random walk among different replicas. Moreover, to improve the performance of the algorithm, the defect is translated randomly around the lattice and local updates are more frequent around it.

Finally, concerning the defect size, we chose it to be approximately constant in physical units, $l_d=aL_d \sim 0.18-0.20$ fm, which was proven to be the optimal choice in previous PTBC studies~\cite{Bonanno:2020hht,Bonanno:2024nba,Bonanno:2025eeb}. Once $l_\dd$ is fixed, the number of replicas is scaled with $b$ and $N$ as $N_r\sim NL_d^{3/2}$ to keep $p\simeq20\%$ constant, see~\cite{Bonanno:2020hht,Bonanno:2024nba,Bonanno:2025eeb} for more details.

\subsection{Improvement of lattice artifacts}\label{sec:improvement}

In order to obtain a tree-level improvement of the lattice observables used to compute \flow and \flowprime, one can resort to perturbation theory. In the continuum and at leading order in the 't Hooft coupling constant~\cite{Ramos:2014kla,GarciaPerez:2014azn,Bribian:2019ybc}:
\be\label{eq:phiprime_pt}
\varphi^{\prime}(t) = \mathcal{N}_{\mathrm{cont}}(N,l,l_s,t) (\lambda + \cdots )
\ee
where:
\be\label{eq:norm_cont}
\mathcal{N}_{\mathrm{cont}}(N,l,l_s,t) = \frac{3}{128\pi^2}\mathcal{A}(x, x_s, N)\, ,
\ee
with
\be
\mathcal{A}(x, x_s, N) = \tilde{\theta}^2\left(x\right) \left[\tilde{\theta}^2\left(x_s\right) - \frac{1}{N^2} \tilde{\theta}^2\left(N^2 x_s\right)\right]
\ee
and
\be
x = \frac{8\pi t}{l^2} \, , \enspace x_s =  \frac{8\pi t}{(Nl_s)^2}\, , \enspace \tilde{\theta}(x) = \theta_3\left(0, \frac{i}{x}\right)\, ,
\ee
with $\theta_3$ the Jacobi theta function.

On the lattice, with the clover definition of the energy density given in \cref{eq:Eclover}, one should replace the normalization $\mathcal{N}_{\mathrm{cont}}$ by the expression obtained when expanding the clover energy density in lattice perturbation theory~\cite{Ramos:2014kla,GarciaPerez:2014azn}: 
\be\label{eq:norm_latt}
\mathcal{N}_{\mathrm{latt}}(N,L,L_s,T) = \frac{T^2}{2 V_{\mathrm{eff}}}\sum_{\mu\,\neq\,\nu,\,q}^{\prime}\frac{e^{2T\hat{q}^2}\sin^2(q_\nu)\cos^2(q_\mu/2)}{\hat{q}^2}
\ee
where $t=T a^2$, $l=L a$, $l_s=L_s a$, $V_{\mathrm{eff}} = (N L_s L)^2$, and $\hat{q}_\mu = 2\sin(q_\mu/2)$ is the lattice momentum, with $q_\mu = 2\pi n_\mu/L_{\mathrm{eff}}$ for $n_\mu = 0,\dots,L_{\mathrm{eff}}-1$. The effective size $L_{\mathrm{eff}}$ is $L$ along the directions $\mu=0,3$ with \PBC and $NL_s$ along  $\mu=1,2$ with \TBC. The prime in the sum denotes the exclusion of momenta with both components on the twisted plane multiple of $2\pi/L_s$, that is  $n_1 \bmod N = 0$ and $n_2 \bmod N = 0$. Thus, a tree-level improvement of lattice artifacts in the flow can be attained by multiplying \flow and \flowprime by
\be\label{eq:improvement-artifacts}
\tilde{\mathcal{N}}(N,L,L_s,T) = \frac{\mathcal{N}_{\mathrm{cont}}(N,l,l_s,t)}{\mathcal{N}_{\mathrm{latt}}(N,L,L_s,T)}\, .
\ee
It should be noted that this choice only removes classical lattice artifacts in the observable used to define the flow, while those associated to the discretization of the lattice action and of the
flow equation are left untouched~\cite{Ramos:2015baa}.

\subsection{Twisted volume reduction in the large-\ensuremath{N} limit}\label{sec:volume-reduction}

One of the objectives of the present paper is to examine how the effects of finite volume are mitigated when using twisted boundary conditions at finite $N$.  One of the consequences of reduction is an enlargement of the effective size in the directions subject to \TBC. For our particular choice of twist, this amounts to replacing $l_s \rightarrow \leff \equiv N l_s$. While this holds exactly in the large-$N$ limit, there is an explicit dependence on the actual torus size at finite $N$, which becomes suppressed in inverse powers of $N$. In order to illustrate this point, consider the leading order in perturbation theory, where the leading volume- and $N$-dependence of the flow can be straightforwardly derived from \cref{eq:phiprime_pt}. Taking into account that
\be
\tilde{\theta}\left(x\right) = \sum_{n \in Z} \exp \left[-\frac{\pi n^2}{x} \right]\, ,
\ee
one gets, at leading order in the coupling:
\be
\varphi^{\prime}(t) = \frac{3 (N^2-1)}{128 \pi^2 N^2}  \left[ 1- \frac{4}{N^2-1} \eee^{-l_s^2/(8t)} \right] \lambda \, ,
\ee
up to exponential corrections in $l^2/(8t)$ and $\left[\leff\right]^2 /(8t)$. This expression reflects the underlying principle of reduction; in the large-$N$ limit, irrespective of the size of the torus in the twisted directions, the thermodynamic limit is recovered with corrections that depend only on the effective volume. The perturbative expansion of Wilson loops has been shown to exhibit a similar behavior, with the dependence on the small size being suppressed as $1/N^2$ in the large-$N$ limit~\cite{Perez:2017jyq,Gonzalez-Arroyo:2014dua}. In accordance with these observations, we conjecture that, by virtue of large-$N$ twisted volume reduction, the finite-size effects on the gradient-flow scales can be adequately characterized as follows:
\be\label{eq:fvol}
\frac{t_0(N,\leff,l)}{t_0(N)} = 1 + G(\leff, l) + \frac{A(N)}{N^2}F\left(\frac{\leff}{N}, l\right)  \, ,
\ee
with $A(N) \sim \mathcal{O}(N^0)$, and $t_0(N) \equiv t_0(N,\infty,\infty)$ the thermodynamic limit of \tzero for a given $N$. Analogous expressions hold for all the other scales. It is important to emphasize that this expression reflects the power of the large-$N$ volume reduction: the term proportional to $F$ tends to zero as $N$ increases and the large-$N$ result coincides with the one for $\leff=\infty$, irrespective of the value of $l_s$.

From the field-theoretical point of view, the actual functional form of the functions $F$ and $G$ depends on whether expectation values are taken in a fixed topological sector, or averaging over all topological sectors. This is crucial to determine the expected finite-size effects in the presence/absence of topological freezing. If topology is correctly sampled and expectation values are taken averaging over all relevant values of $Q$, finite-size effects on the energy density are expected to be exponentially suppressed in $l$, $l_s$ and $\leff$, at least if these sizes are large enough in units of $\sqrt{8 t_0}$ to be sufficiently deep in the non-perturbative domain in which \tzero is defined. Given that $\leff,l>l_s$, finite-size corrections at finite $N$ will be dominantly due to the $F$-term and to its dependence on the short size:
\be\label{eq:FSE_noproj}
\frac{t_0(N,\leff,l)}{t_0(N)} \simeq 1 + \frac{A_0}{N^2}\eee^{-m \leff/N} + \dots \, ,
\ee
with $m$ some physical mass scale of the theory. Instead, if the gradient-flow scales are computed in a fixed topological sector, finite-size effects in $l$, $l_s$ and $\leff$ are expected to be powerlike, and $\mathcal{O}(1/\Vc)$ (for $F$) and $\mathcal{O}(1/\Veffc)$ (for $G$) at leading order~\cite{Brower:2003yx,Aoki:2007ka}, with $\Veffc \equiv N^2 \Vc= N^2 l^2 l_s^2$. However, due to the $1/N^2$ suppression of the $F$-term with respect to $G$, eventually also that term becomes $\mathcal{O}(1/\Veffc)$. Thus, in a fixed topological sector $Q=n$, the dominant finite-size correction will be:
\be\label{eq:FSE_proj}
\left.\frac{t_0(N,\leff,l)}{t_0(N)}\right\vert_{Q\,=\,n} \simeq 1 + \frac{C^{(n)}(N)}{N^2\Vc} + \dots\\
\ee
with $C^{(n)}(N)\sim\mathcal{O}(N^0)$.

\section{Results}\label{sec:results}

In this section, we present our main results for the gradient-flow scales. After a brief explanation of our choice of the simulation parameters, we present the determination of the scales on an $N$-by-$N$ basis and analyze the systematic effects associated to finite volume and topological freezing. Next, we study the dependence on the number of colors and perform a global analysis of finite-volume and finite-$N$ effects to illustrate the effect of reduction. We finally present results on the continuum extrapolation of various ratios of scales to analyze the effect of the improvement of lattice artifacts. 

\subsection{Simulation parameters}\label{sec:results:simulations}

The simulated values of the inverse bare 't Hooft coupling $b$ selected for each value of $N$ in this study were determined based on the requirements for computing the 
corresponding \SUN $\Lambda$ parameter via step-scaling. To this end, a finite-volume renormalization scheme is employed, whereby the running coupling scale is set proportional to the box size: $\mu\propto l^{-1}$.  In this particular context, the coupling is evolved from a conventional hadronic reference scale $\muhad$ into the perturbative domain. In this domain, perturbation theory is applicable, and a a matching to a perturbatively defined scheme, such as $\overline{\rm MS}$, can be performed. The reference scale $\muhad$ should then be converted to physical units using a standard reference scale, such as $t_0$, by determining $\muhad \times \sqrt{t_0}$, in the continuum limit. Scale setting enters exactly here: to determine $\muhad$ in units of $\sqrt{t_0}$, one needs to know $\sqrt{t_0}/a$ in the range of lattice spacings where $a \muhad $ is defined.

In our step-scaling study, the reference scale $\muhad$ is defined by the condition that the renormalized 't Hooft coupling in the Twisted Gradient Flow (TGF) scheme, as defined in Ref.~\cite{Bribian:2021cmg}, is set to a given value $\lambda(\muhad)$, chosen in such a way that a determination of $\sqrt{t_0}/a$ is feasible while keeping lattice artifacts under control. The TGF renormalization scheme is defined in a \TBC setup analogous to the one presented in this paper, the only difference being that the aspect ratio $l/l_s$ is maintained fixed and equal to the number of colors $N$ (for further details see Ref.~\cite{Bribian:2021cmg}). The TGF coupling $\lambda_{\TGF}^{(0)}(\mu)$ is defined as
\be
\lambda_{\TGF}^{(0)}\left(\mu\right)  =  \left.\mathcal{N}_{\mathrm{cont}}^{-1}\left (N,l,l/N,t\right ) \, \varphi^{\prime}_0(t) \right\vert_{\sqrt{8 t} \,=\, c l \,=\, \mu^{-1}}  \, ,
\label{eq:lambdat}
\ee
where $t$ is the flow time and the renormalization scale $\mu$ is determined by relating the smoothing radius $\sqrt{8 t}$ of the flow to a fraction ($c=0.3$) of the box size $l$. This scheme has several advantages, such as a reduced memory footprint with respect to standard lattice simulations on symmetric ($l=l_s$) lattices.

When it comes to tuning the lattices that define $a\muhad$, our choices for  lattice sizes $L_i$ are constrained by two requirements: $L_i$ must be a multiple of $N$ to have an integer $L_s$, and $a(b_i)\lesssim 0.1$ fm to avoid excessively large lattice artifacts. This has led to select two particular choices of the TGF coupling, $\lambda_{\TGF}^{(0)}(\muhad) \simeq 34.43$ ($N=3$) and $44.81$ $(N=5,8)$. These correspond to constant lattice sizes $l=1.1$ and 1.3 fm (up to lattice artifacts) respectively. For each lattice size $L_i$, the value of the bare coupling $b_i$ corresponding to a given renormalized coupling is determined following the strategy described in Refs.~\cite{Bribian:2021cmg,Bonanno:2024nba}. The process involves the fitting of the dependence of the renormalized coupling on $b$ at every given value of $L$. The results of the tuning procedure are reported in \cref{tab:tuned_simulations}. It is important to stress that this choice implies a rather rapid decrease in the required lattice spacings. As an example, for $N=5$, the choices $L_i=20,30,40$ and $l\simeq 1.3$ fm imply that $a(b_i)\simeq 0.065,\ 0.043,\ 0.033$ fm respectively.

\begin{table}[!t]
\centering
\begin{tabular}{cccc}
\toprule
$N$ & $b$ & $L$ & $L_s$ \\
\midrule
\multirow{3}{*}[-0.0em]{3}  &    0.35883  & 24 &  8 \\
&    0.37583  & 36 & 12 \\
&    0.38844  & 48 & 16 \\
\bottomrule
\end{tabular} \,
\begin{tabular}{cccc}
\toprule
$N$ & $b$ & $L$ & $L_s$ \\
\midrule
\multirow{3}{*}[-0.0em]{5}  &    0.35971  & 20 &  4 \\
&    0.37504  & 30 &  6 \\
&    0.38683  & 40 &  8 \\
\bottomrule
\end{tabular} \,
\begin{tabular}{cccc}
\toprule
$N$ & $b$ & $L$ & $L_s$ \\
\midrule
\multirow{3}{*}[-0.0em]{8}  &    0.35867  & 16 &  2 \\
&    0.38352  & 32 &  4 \\
&    0.40008  & 48 &  6 \\
\bottomrule
\end{tabular}
\caption{Simulation parameters of the lattices tuned to have a TGF coupling $\lambda_{\TGF}^{(0)}(\muhad) \simeq 44.81$ ($N=5,8$) and $\lambda_{\TGF}^{(0)}(\muhad) \simeq 34.43$ ($N=3$). They correspond respectively to $l\simeq 1.3$ fm and $1.1$ fm, where $l = 1/(0.3 \,\muhad)$.}
\label{tab:tuned_simulations}
\end{table}

\begin{figure}[!t]
\centering
\includegraphics[scale=0.35]{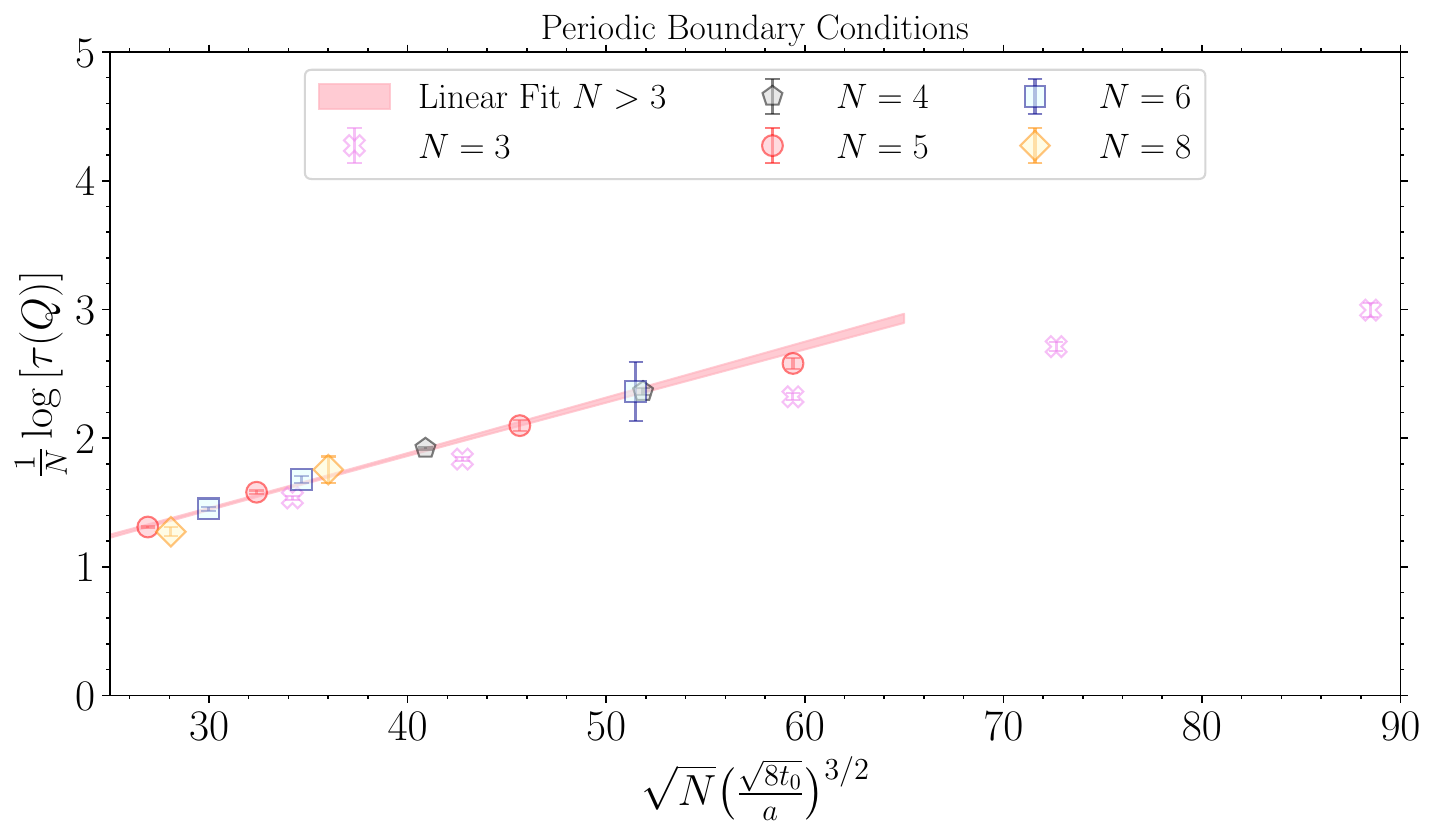}
\caption{Approximate collapse of $\log[\tau(Q)]/N$ ($N>3$) as a function of $\sqrt{N}/a^{3/2}$ with \PBC. This plot uses data from Ref.~\cite{Athenodorou:2021qvs}. Lattice spacing determinations of~\cite{Athenodorou:2021qvs} in terms of the string tension $\sigma$ have been converted to $t_0$ units using the result $\sqrt{8t_0\sigma} = 1.2068(46) - 0.997(69)/N^2$ of~\cite{Bonanno:2025eeb}.}
\label{fig:tau_PBC_collapse}
\end{figure}

It is clear that, for scale-setting purposes, all tuned bare couplings fall within the regime where topological freezing is very severe. As an example, in the case of $N=5$ discussed above, we estimate autocorrelation times for PBCs as $\tau(Q)\simeq 2.8 \cdot 10^{4}, 8.5 \cdot 10^{7}, 4.0 \cdot 10^{11}$, for each of the tuned lattice spacings. These values are derived using the results for $\tau(Q)$ obtained in Ref.~\cite{Athenodorou:2021qvs}, by noting that the $\log[\tau(Q)]/N$ data ($N>3$) follow a common linear curve when plotted against the variable $\sqrt{N}/a^{3/2}$, see \cref{fig:tau_PBC_collapse}. Thus, performing the scale setting in this regime using standard methods would be impossible. On the other hand, with PTBC the autocorrelation times of $Q$ in the periodic replica (at fixed acceptance and defect size in physical units) scale polynomially with both $N$ and $1/a$: $\tau_0(Q) \sim \sqrt{N}/a^2$. The required number of replicas also exhibits a polynomial scaling $N_r \sim N/a^{3/2}$. These behaviors were demonstrated in~\cite{Bonanno:2020hht,Bonanno:2024nba,Bonanno:2025eeb}. As discussed in Ref.~\cite{Bonanno:2025eeb}, this ultimately leads PTBC to outperform both \PBC and \OBC simulations at fixed numerical effort. As an example, using the data reported in Ref.~\cite{Ce:2016awn} for \OBC and in Ref.~\cite{Bonanno:2025eeb} for PTBC, one can estimate $\tau(Q)\simeq 5\cdot10^4$ and $\tau_0(Q)\simeq 5\cdot10^2$ (with 20 replicas) lattice sweeps, respectively, for the coarsest lattice spacing of $N=5$ considered here~\footnote{This comparison was performed by normalizing the cost of 1 lattice sweep to the same lattice volume in order to take into account the fact that \OBC simulations need a larger time extent to avoid boundary effects. See Ref.~\cite{Bonanno:2025eeb} for a thorough discussion on this point.}.

\cref{tab:summary_simulations} provides a comprehensive overview of all the simulation parameters used in this study. Even when using PTBC, one has to face a nontrivial computational challenge to explore the required regime. Given that we only need scale setting determinations at the bare couplings in \cref{tab:tuned_simulations}, the most efficient approach is to perform dedicated simulations at these values. In addition, we combined PTBC with finite-size scaling to minimize the memory requirements of our computation. For each combination of $N$ and $b$, simulations were performed for several lattice volumes, considering both projected and non-projected gradient-flow scales in order to achieve full control over the shape of finite-size effects, according to \cref{eq:FSE_noproj} and \cref{eq:FSE_proj}. This allows to subtract finite-size effects in both the short and the long size and to achieve solid determinations in the infinite-volume limit, as well as to check the agreement of projected and non-projected quantities when $l\to\infty$. This approach is essential for controlling finite-volume effects, even in cases where performing direct simulations on sufficiently large volumes would require significant time and memory resources (to store all the replicas for large lattice volumes). Indeed, our code implementation for PTBC can only run on a single-node setup. While this problem could in principle be avoided by parallelizing the code across several nodes, in practice this would require rewriting our code from scratch. This limitation affects our simulations on the finest lattice spacing of $N=5$ and $8$. In these cases, we opted to take the infinite-volume limit of scale determinations obtained from standard PBCs simulations (frozen in $Q = 0$). We anticipate that, in all other cases where explicit PTBC simulations were performed, the infinite-volume limit of projected data will always yield results that agree perfectly with the infinite-volume limit of unfrozen PTBC determinations (as expected from general theoretical arguments). As detailed in Appendix~\ref{sec:appendix}, this approach has been further crosschecked by performing one explicit $N=5$ PTBC simulation on the finest lattice for one intermediate value of the volume, and then subtracting the finite-size effects determined from coarser ensembles. This resulted in perfect agreement with the infinite-volume limit of frozen data, thus this further point has been included in the analysis too.

Finally, at the end of our calculation, an important cross-check will be to explicitly verify that bare couplings tuned to correspond to the same renormalized $\lambda_{\TGF}^{(0)}(\muhad)$ also correspond to the same value of the lattice size $l$ in units of some gradient-flow scale.

\begin{table}[!t]
\centering
\begin{tabular}{ccccccccc}
\toprule
$N$ & $b$ & $\beta$& Algorithm & $N_r$ & $L_d$ & $V_{\rm max}$ & $\Delta_s$  & $N_s$ \\
\midrule
\multirow{3}{*}[-0.0em]{3} &    0.35883  &  6.459 &    PTBC  &    18  & 4 & $24^236^2$ &   4 & $1.4\cdot10^4$ \\
                           &    0.37583  &  6.765 &    PTBC  &    34  & 6 & $30^254^2$ &  12 & $1.6\cdot10^3$ \\
                           &    0.38844  &  6.992 &    PTBC  &    54  & 8 & $32^260^2$ &  16 & $5.7\cdot10^2$ \\
\addlinespace[0.5em]
\multirow{4}{*}[-0.0em]{5} &    0.35971  & 17.985 &     PTBC  &    21  &  3  & $16^226^2$ &   6 & $1.6\cdot10^4$ \\
                           &    0.37504  & 18.752 &     PTBC  &    32  &  4  & $22^240^2$ &  8 & $3.9\cdot10^3$ \\
                           &    \multirow{2}{*}[-0.0em]{0.38683}  & \multirow{2}{*}[-0.0em]{19.342} &     PTBC &    44  &  5  & $16^242^2$ &  12 & $2.5\cdot10^3$ \\
                           &                                      &                                 &     Std. &     1  & --  & $28^246^2$ & 120 & $3.2\cdot10^3$ \\
\addlinespace[0.5em]
\multirow{3}{*}[-0.0em]{8} &    0.35867  & 45.910 &     PTBC  &    18  &   2  & $12^218^2$ &   4 & $2.5\cdot10^4$ \\
                           &    0.38352  & 49.091 &     PTBC  &    46  &   4  & $24^236^2$ &  8 & $1.4\cdot10^3$ \\
                           &    0.40008  & 51.210 &     Std.  &     1  &  --  & $36^260^2$ & 80 & $6.1\cdot10^2$ \\
\bottomrule
\end{tabular}
\caption{Summary of simulation parameters using \TBC. We report both the inverse bare 't Hooft coupling $b=(N^2 g)^{-1}$ and the standard inverse bare gauge coupling $\beta=2N^2b$. With the Standard algorithm, only the physical replica is simulated and there is no defect. In every case, the number of over-relaxation sweeps per heat-bath sweeps is $n_{\ov} = 12$. $V_{\rm max}$ is the largest volume simulated for each ($N$, $b$). For the PTBC algorithm, the number $\Delta_s$ of Monte Carlo steps between measures was chosen such that the integrated autocorrelation time of $E(t\simeq t_0)$ is about one. The reported statistics $N_s$ is the number of measures after thermalization, averaged over all the volumes simulated for each ($N$, $b$).}
\label{tab:summary_simulations}
\end{table}

\subsection{Scales on an \ensuremath{N}-by-\ensuremath{N} basis}\label{sec:results:volume-NbyN}

We determined the gradient-flow scales with \TBC for three lattice spacings for each value of $N=3,5,8$, with several combinations of lattice sizes. We used both \cref{eq:phi} and \cref{eq:phi_prime} as definitions of the flow, with and without the improvement in \cref{eq:improvement-artifacts}. We report the full raw results in the Supplemental Material~\cite{suppmat}.

The following analysis of finite-volume effects was carried over in the same way for each $N$ separately and for all the definitions of the scales, only distinguishing whether using the topological projection to \qzero or not. Thus, we call a generic scale in lattice units $S$ if not projected and $S^{(0)}$ if projected. For the sake of maintaining the discussion compact, we will only show results for \tzero with the normalization of \cref{eq:phi}.

Let us start from the scales obtained considering all topological sectors. We will assume that finite-volume corrections are independent of the lattice spacing, once the lattice sizes are expressed in physical units. This allows to fit data obtained at different values of $b$ at once. Also, since at this stage we are analyzing each $N$ separately, we are not enforcing any particular $N$-dependence in the fit parameters. This will allow to check the expected $1/N^2$ suppression of finite-size corrections in the short size $l_s$.

Concerning possible finite-volume corrections in the long size $l$ and the effective size $\leff = Nl_s$, we expect them to be exponentially suppressed as those in $l_s$, as discussed in \cref{sec:volume-reduction}. Given that $\leff > l > l_s$, we will assume them to be subleading and negligible. Thus, for each combination of $N$, $b$ and $l_s$, we just took the largest $l = l_{\mathrm{max}}$ among those simulated and fitted only the following dependence on $l_s$:
\be\label{eq:fit-volume-NbyN-anyQ}
\frac{S(N, b, L_s,L_{\mathrm{max}})}{S(N, b)} = 1 - \hat{A}(N) \, \eee^{-M(N)\,\hat{L}_s(N,b)}\, .
\ee
with $\hat{L}_s(N,b)=L_s/\sqrt{8S(N,b)}$. The underlying assumption, which we will check \emph{a posteriori} later, is that the values of $l_{\mathrm{max}}$ entering the fitted dataset are large enough in physical units to neglect their associated volume corrections at our statistical precision. In particular, we anticipate that our data will satisfy $\hat{L}_{\mathrm{max}}(N,b) \equiv L_{\mathrm{max}}/\sqrt{8S(N,b)} \gtrsim 3$ in every case, a value which has been shown in previous studies to be sufficient to contain finite-size effects (see, e.g., Refs.~\cite{Ce:2015qha,Ce:2016awn}).

\begin{figure}[!t]
\centering
\includegraphics[width=\columnwidth]{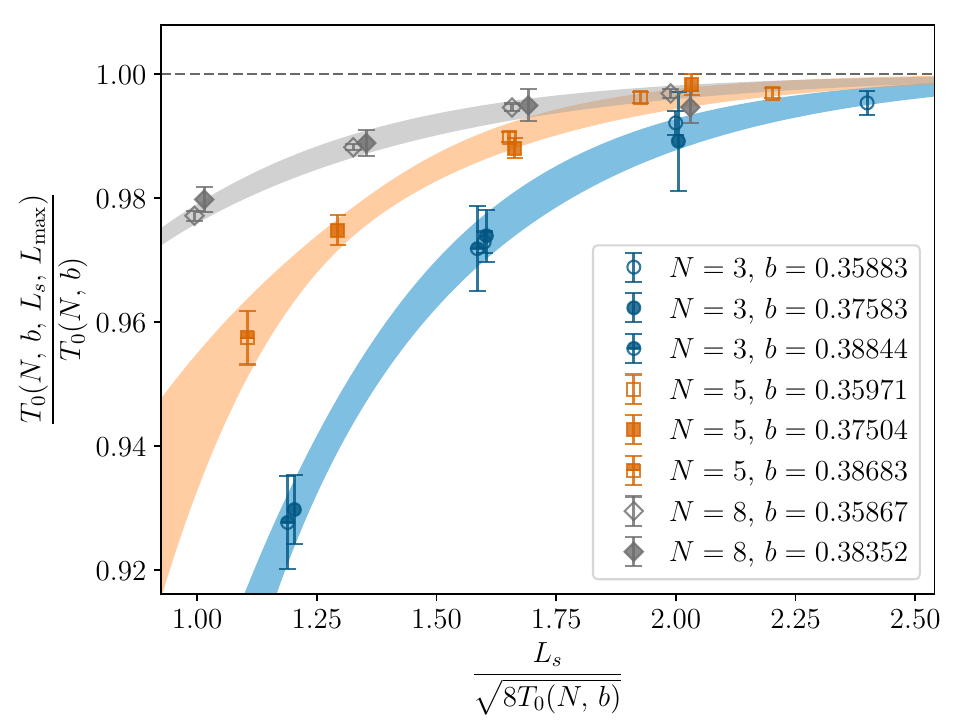}
\caption{Dependence on the short size $\hat l_s$ of the non-projected scale \tzero with improvement of lattice artifacts, see \cref{eq:improvement-artifacts}. The scales $T_0(N,b,L_s, L_{\mathrm{max}})$ are determined at the largest long size $l=l_{\mathrm{max}}$ available, and satisfy $l_{\max}\gtrsim 3\sqrt{8t_0}$. All finite-volume scales are normalized with their corresponding infinite-volume result $T_0(N,b)$ obtained from the $N$-by-$N$ fits.}
\label{fig:t0+_NbyN_twst_imp2_anyQ_vs_Ls}
\end{figure}

\begin{figure}[!t]
\centering
\includegraphics[width=\columnwidth]{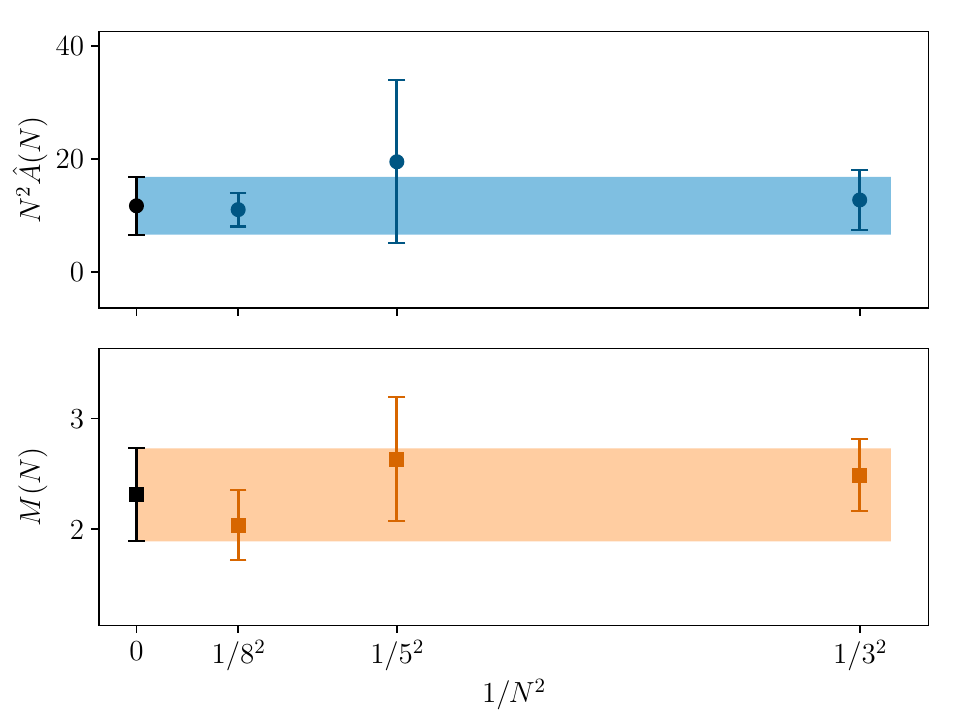}
\caption{Dependence on $N$ of the best-fit parameters $\hat{A}(N)$ and $M(N)$ in \cref{eq:fit-volume-NbyN-anyQ}, obtained from the $N$-by-$N$ fits of the improved \tzero shown in \cref{fig:t0+_NbyN_twst_imp2_anyQ_vs_Ls}. As expected, $\hat{A}(N) = \mathcal{O}(N^{-2})$ and $M(N) = \mathcal{O}(N^{0})$.}
\label{fig:t0+_NbyN_twst_imp2_anyQ_params_vs_N}
\end{figure}

In \cref{fig:t0+_NbyN_twst_imp2_anyQ_vs_Ls} we show the fitted $\hat{l}_s$-dependence for the improved \tzero. To show all the values of $N$ and $b$ in the same figure, we plot the ratio $T_0(N,b,L_s,L_{\mathrm{max}})/T_0(N,b)$, where the numerator is our lattice determination, while the denominator is the infinite-$l_s$ limit determined from the fits. Results for different lattice spacings fall on the same curve, confirming our hypothesis that IR finite-size effects are practically insensitive to UV discretization effects.

As it can be seen in \cref{fig:t0+_NbyN_twst_imp2_anyQ_vs_Ls}, finite-size effects in the short size are well-described by an exponential suppression, and become smaller at larger $N$, signaling a suppression of the prefactor $\hat{A}(N)$ in the large-$N$ limit. Such suppression is expected to be $\sim1/N^2$, see \cref{eq:fvol}. In \cref{fig:t0+_NbyN_twst_imp2_anyQ_params_vs_N} we indeed see that $N^2 \hat{A}(N)$ has a finite large-$N$ limit, in agreement with our expectations. In the same figure we also show the exponential decay constant $M$, which turns out to not depend significantly on $N$, and to be of the same order of magnitude of the Yang--Mills mass gap. Indeed, for the large-$N$ limit of $M=\sqrt{8t_0} m$ we find:
\be
M_{\scriptscriptstyle{\infty}} = \sqrt{8t_0}\minf = \mathrm{2.31(42)} \, .
\ee
Using $\sqrt{8t_0\sigma}\big\vert_{N\,=\,\infty}=1.207(5)$~\cite{Bonanno:2025eeb}, we find $\minf/\sqrt{\sigma}=\mathrm{1.91(34)}$, to be compared with the large-$N$ Yang--Mills mass gap $M_{\scriptscriptstyle{\rm G}}^{\scriptscriptstyle{(\infty)}}/\sqrt{\sigma}=3.072(14)$~\cite{Athenodorou:2021qvs}.

\begin{figure}[!t]
\centering
\includegraphics[width=0.95\columnwidth]{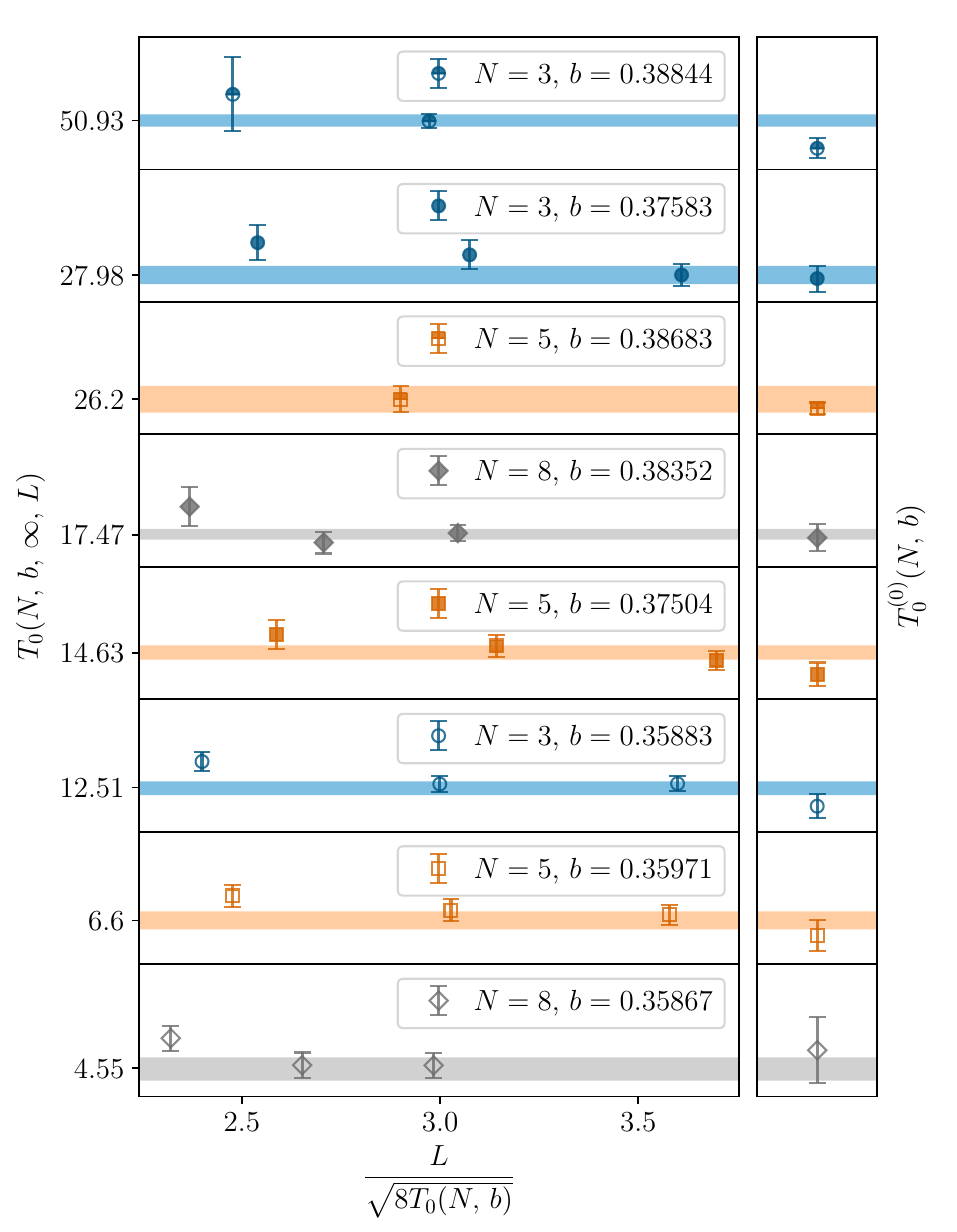}
\caption{Left panels: dependence of the non-projected scale \tzero on the long size $l$ after subtraction of the exponentially-suppressed corrections in $l_s$. Right panels: infinite-volume results for the projected scale \tzeroproj. Colored bands represent the infinite-volume result for the non-projected scale.}
\label{fig:t0+_NbyN_twst_imp_all_vs_L}
\end{figure}

To check whether the exponential corrections in the long size are negligible as we assumed, we subtracted the fitted exponential corrections in the short size from the determination of the non-projected scales obtained at different values of $l$ and at one intermediate value of $l_s$. Results for the improved \tzero, showed in the left panels of \cref{fig:t0+_NbyN_twst_imp_all_vs_L}, show no residual dependence on $\hat{l}$ for $l \gtrsim 2.7\sqrt{8t_0}$. In this case, all results are compatible among them and with the infinite-volume determination, represented by shaded bands in the figure.

\begin{figure}[!t]
\centering
\includegraphics[width=0.98\columnwidth]{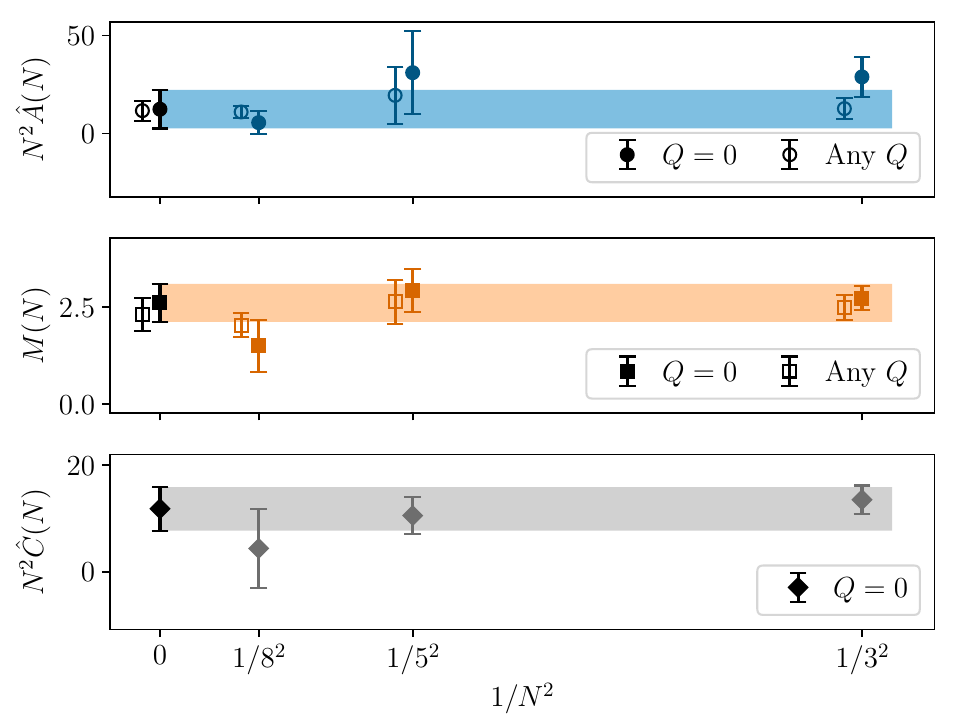}
\caption{Dependence on $N$ of the best-fit parameters $\hat{A}(N)$, $M(N)$ and $\hat{C}(N)$ in \cref{eq:fit-volume-NbyN-Q=0}, obtained from the $N$-by-$N$ fits of the improved \tzeroproj. The values resulting from the non-projected fits are also shown. Also in this case, $\hat{A}(N) = \mathcal{O}(N^{-2})$ and $M(N) = \mathcal{O}(N^{0})$, and values obtained with and without projection are compatible at each $N$. Moreover, also $\hat{C}(N) = \mathcal{O}(N^{-2})$ holds as expected.}
\label{fig:t0+_NbyN_twst_imp2_Q=0_params_vs_N}
\end{figure}

\begin{figure}[!t]
\centering
\includegraphics[width=0.95\columnwidth]{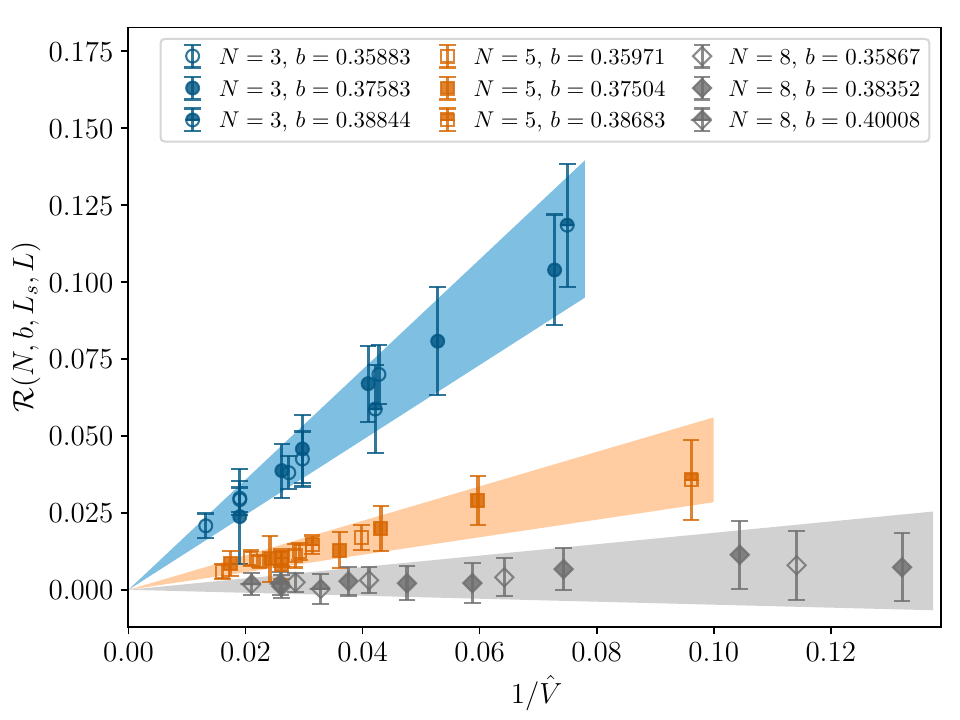}
\caption{Residual dependence of the projected scale $t_0^{(0)}$ on the volume $\hat{V} = \hat{L}^2\hat{L}_s^2$ in physical units, after subtraction of the fitted subleading exponential corrections in the short size. The quantity $\mathcal{R}$ is defined in \cref{eq:residual-fit-correction-NbyN}.}
\label{fig:t0+_NbyN_twst_imp2_Q=0_vs_V}
\end{figure}

Let us now discuss the scales obtained after projection to $Q=0$. Again, we will neglect exponentially-suppressed corrections in the long sizes $l$, $\leff$, and we will assume finite-size effects to be the same for all $b$ at fixed $N$. Concerning the dependence on $l_s$, the leading term appears through the topological correction, which scales as $1/(l \, \leff)^2 = 1/(N \,l\,l_s)^2$, but in our fit we will also consider the subleading exponentially-suppressed term in $l_s$. It will turn out to be much smaller compared to the powerlike one, but nonetheless present. Finally, we again fit each data set at different values of $N$ independently and without enforcing any $N$-dependence in the prefactor. In the end, we use the following fit function:
\be\label{eq:fit-volume-NbyN-Q=0}
\frac{S^{(0)}(N,b,L_s,L)}{S^{(0)}(N,b)} = 1 + \frac{\hat{C}(N)}{\hat{V}} - \hat{A}(N)\, \eee^{-M(N)\hat{L}_s(N,b)} \, ,
\ee
with $\hat{V} = \hat{L}^2 \hat{L}_s^2$ and $\hat{L}_s(N,b)=L_s/\sqrt{8S^{(0)}(N,b)}$. As for the fit without projection, we are neglecting exponentially-suppressed corrections in the long sizes $l$ and $\leff$. We only include in the fit those points with $\hat{L} \gtrsim 2.7$, for which we do not appreciate finite-$\hat{L}$ effects in the non-projected scales. For \tzeroproj, we show in \cref{fig:t0+_NbyN_twst_imp2_Q=0_params_vs_N} that the prefactor $\hat{C}(N)$ is suppressed as $1/N^2$, and that the exponential decay constant $M(N)$ and its prefactor $\hat{A}(N)$ turn out to be in agreement with the ones found for the non-projected data. These results also hold for the other scale definitions. In \cref{fig:t0+_NbyN_twst_imp2_Q=0_vs_V}, in order to show the finite-size effects associated with the $Q=0$ projection, we plot the quantity
\be\label{eq:residual-fit-correction-NbyN}
\begin{aligned}
\mathcal{R}(N, b, L_s, L ) = & \, \frac{T^{(0)}_0(N, b, L_s, L)}{T^{(0)}_0(N, b)} - 1 \, + \\
+ & \, \hat{A}(N)\, \eee^{-M(N)\hat{L}_s(N,b)} \, ,
\end{aligned}
\ee
that is, the residual finite-size corrections after the subtraction of the fitted exponential term, which is compatible with the one of the non-projected scale \tzero. It can be seen that finite-size corrections after $Q=0$ projection are indeed dominated by powerlike terms, which are suppressed as $N$ is increased.

\begin{table}[!t]
\centering
\begin{tabular}{ccccc}
\toprule
\multicolumn{5}{c}{Improved --- $N$-by-$N$ fit strategy} \\
\midrule
$N$ & $b$ & $t_0/a^2$  & $t_1/a^2$  & $w_0^2/a^2$ \\
\midrule
\multirow{3}{*}[-0.0em]{3} &    0.35883  &    12.483(36)[22]  &    5.2414(75)[41]  &    12.993(75)[44] \\
                           &    0.37583  &      27.96(14)[1]  &     11.734(34)[1]  &      29.11(25)[2] \\
                           &    0.38844  &     50.47(36)[63]  &     21.11(10)[10]  &      52.7(6)[1.1] \\
\addlinespace[0.5em]
\multirow{3}{*}[-0.0em]{5} &    0.35971  &      6.595(12)[3]  &    2.6794(22)[19]  &      7.115(38)[2] \\
                           &    0.37504  &    14.606(30)[26]  &      5.939(6)[10]  &    15.784(87)[22] \\
                           &    0.38683  &    26.067(97)[19]  &     10.590(22)[5]  &      28.21(23)[2] \\
\addlinespace[0.5em]
\multirow{3}{*}[-0.0em]{8} &    0.35867  &    4.5507(73)[13]  &     1.8193(10)[4]  &      4.987(22)[1] \\
                           &    0.38352  &     17.470(34)[1]  &     7.0045(62)[4]  &    19.182(92)[11] \\
                           &    0.40008  &      39.16(16)[1]  &     15.728(25)[1]  &      42.86(35)[7] \\
\bottomrule
\end{tabular}
\caption{Infinite-volume scale-setting results of the $N$-by-$N$ analysis with \TBC, flow normalized as in \cref{eq:phi} and improvement of lattice artifacts in \cref{eq:improvement-artifacts}. Results with and without projection onto the \qzero topological sector are combined into a single value with a systematic error (in square brackets), as explained in the text.}
\label{tab:scale-setting-comb-therm-NbyN-twst-imp2}
\end{table}

As a further consistency check, we compared projected scales in the thermodynamic limit, displayed in the right panels of \cref{fig:t0+_NbyN_twst_imp_all_vs_L}, with the infinite-volume results obtained from the non-projected ones. As it can be seen, we find good agreement among the two determinations. The largest difference that we observe is about two standard deviations for the finest lattice spacing at $N=3$, which is likely due to an under-sampling of topological fluctuations due to insufficient statistics.

\begin{table}[!t]
\centering
\begin{tabular}{ccccc}
\toprule
\multicolumn{5}{c}{Unimproved --- $N$-by-$N$ fit strategy} \\
\midrule
$N$ & $b$ & $t_0/a^2$  & $t_1/a^2$  & $w_0^2/a^2$ \\
\midrule
\multirow{3}{*}[-0.0em]{3} &    0.35883  &    12.528(35)[23]  &    5.3052(76)[43]  &    13.000(75)[44] \\
                           &    0.37583  &      28.00(14)[1]  &     11.797(34)[1]  &      29.12(25)[2] \\
                           &    0.38844  &     50.49(36)[63]  &     21.17(10)[10]  &      52.7(6)[1.1] \\
\addlinespace[0.5em]
\multirow{3}{*}[-0.0em]{5} &    0.35971  &      6.643(11)[4]  &    2.7463(23)[21]  &      7.120(39)[2] \\
                           &    0.37504  &    14.651(29)[27]  &      6.005(6)[10]  &    15.790(88)[21] \\
                           &    0.38683  &    26.104(96)[22]  &     10.655(22)[4]  &      28.23(23)[2] \\
\addlinespace[0.5em]
\multirow{3}{*}[-0.0em]{8} &    0.35867  &    4.6020(73)[13]  &     1.8883(10)[4]  &      4.986(23)[1] \\
                           &    0.38352  &     17.518(33)[1]  &     7.0691(62)[2]  &    19.184(95)[13] \\
                           &    0.40008  &      39.21(16)[1]  &     15.795(25)[1]  &      42.86(35)[7] \\
\bottomrule
\end{tabular}
\caption{Infinite-volume scale-setting results of the $N$-by-$N$ analysis with \TBC, flow normalized as in \cref{eq:phi} and no improvement of lattice artifacts. Results with and without projection onto the \qzero topological sector are combined into a single value with a systematic error (in square brackets), as explained in the text.}
\label{tab:scale-setting-comb-therm-NbyN-twst-std}
\end{table}

In order to give a final result for the scales, we combined the projected and non-projected extrapolated results into one determination, and assigned to each scale a statistical and a systematic error based on the differences between the two determination:
\begin{enumerate}
\item For the central value, we took the average between the projected and the non-projected infinite-volume scales, weighted with their statistical errors.
\item For the statistical error, we just took the smallest of the two.
\item For the systematic error, we took the semi-difference between the projected and the non-projected scales, weighted with an estimate of the probability that this difference is not due to a statistical fluctuation:
\be\label{eq:syst}
\Delta_{\rm syst} = \frac{\vert S-S^{(0)} \vert}{2}  \, \mathrm{erf}\left(\frac{\vert S-S^{(0)} \vert}{2\sqrt{2}\Delta_{\rm stat}^{\rm (comb)}}\right)\, ,
\ee
with
\be
\Delta_{\rm stat}^{\rm (comb)} = \sqrt{\Delta^2_{\rm stat}S + \Delta^2_{\rm stat}S^{(0)}} \, .
\ee
\end{enumerate}

\begin{figure}[!t]
\centering
\hspace{-25pt}\includegraphics[scale=0.51]{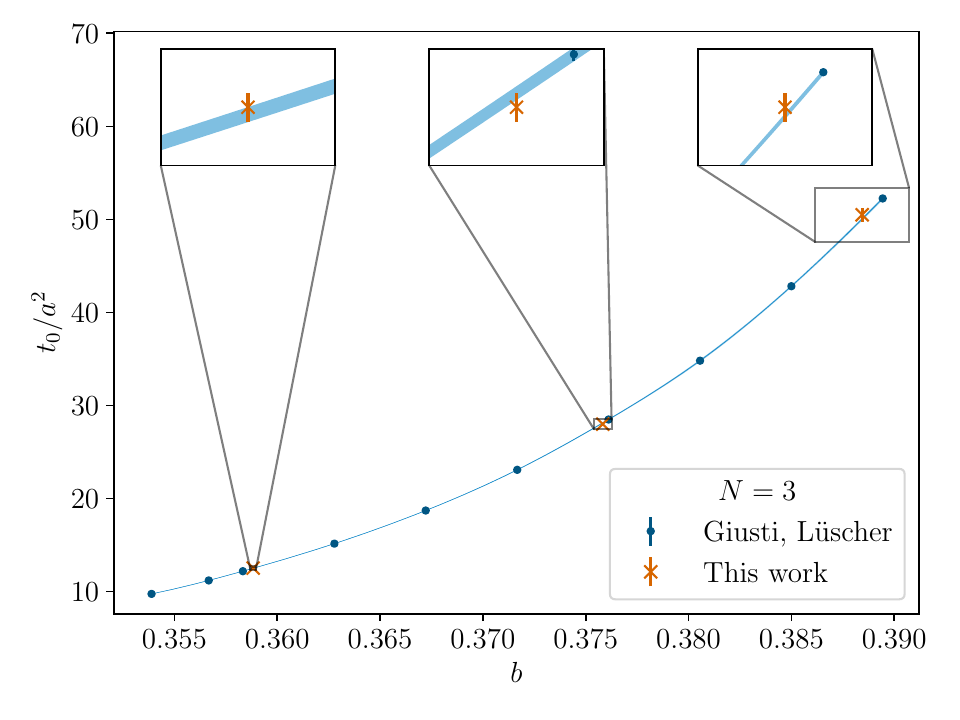}
\includegraphics[width=\columnwidth]{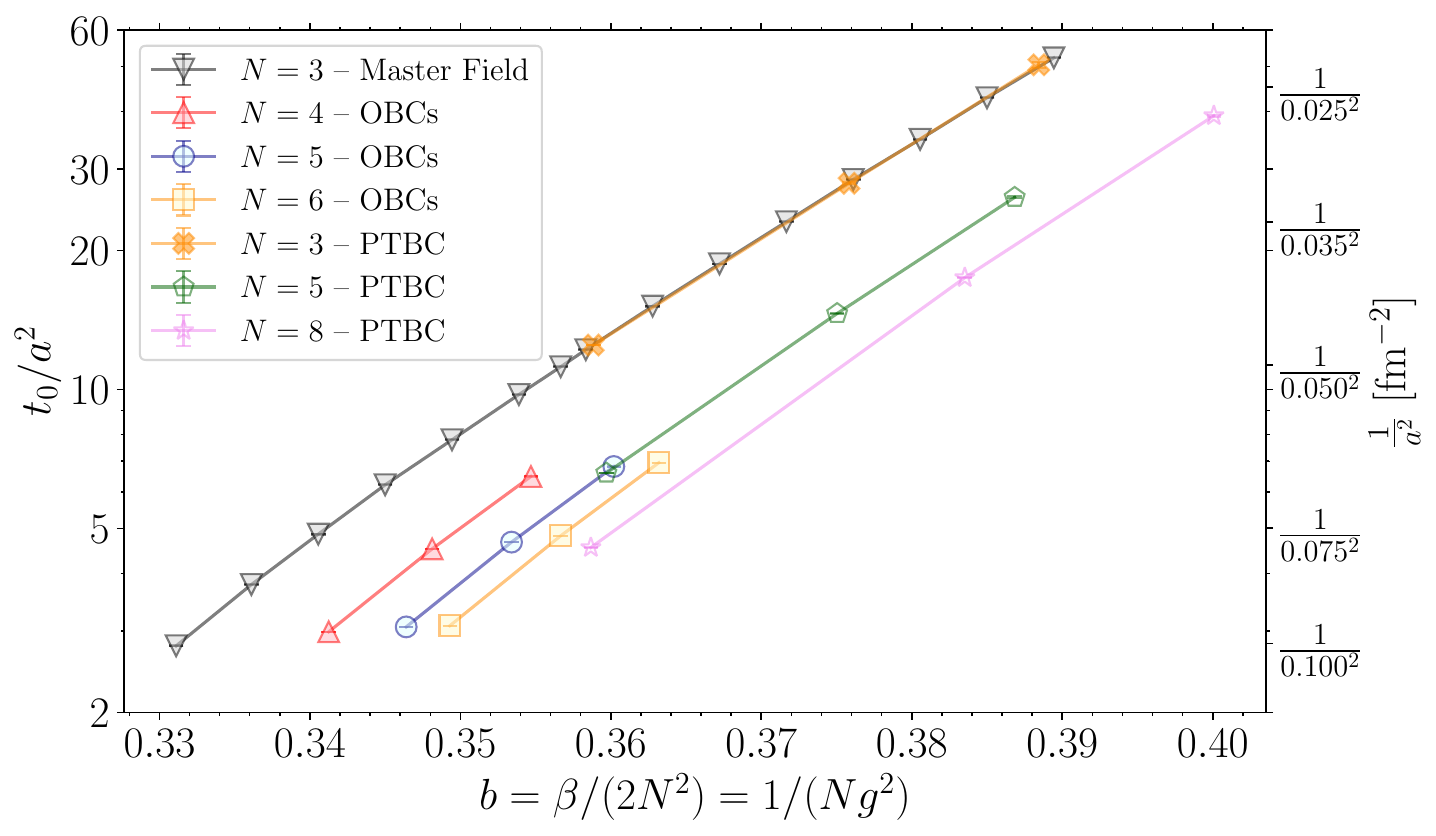}
\caption{Top panel: detailed comparison of our results for the \SU{3} unimproved scale \tzero with a spline interpolation of results of Ref.~\cite{Giusti:2018cmp} (shaded band). Displayed error bars on our points are the quadrature sum of statistical and systematic uncertainties. Bottom panel: summary of present and previous large-$N$ scale setting results for $t_0/a^2$, cf.\cref{fig:t0_summary}.}
\label{t0+_NbyN_twst_std_comb_vs_beta}
\end{figure}

For the finest lattice spacings of \SU{8}, we only have the scales projected into $Q=0$ at our disposal. In these cases, we assigned as a systematic error a fraction of the finite-volume powerlike correction due to the topological projection evaluated on the largest volume available: $\Delta_{\rm syst} = \hat{C}(N)/(4\hat{V})$. We checked that this gives a good upper bound on the systematic error we took for the other cases via \cref{eq:syst}. Final results for the improved scales, cf. \cref{eq:phi}, are reported in \cref{tab:scale-setting-comb-therm-NbyN-twst-imp2}.

\begin{figure}[!t]
\centering
\includegraphics[scale=0.34]{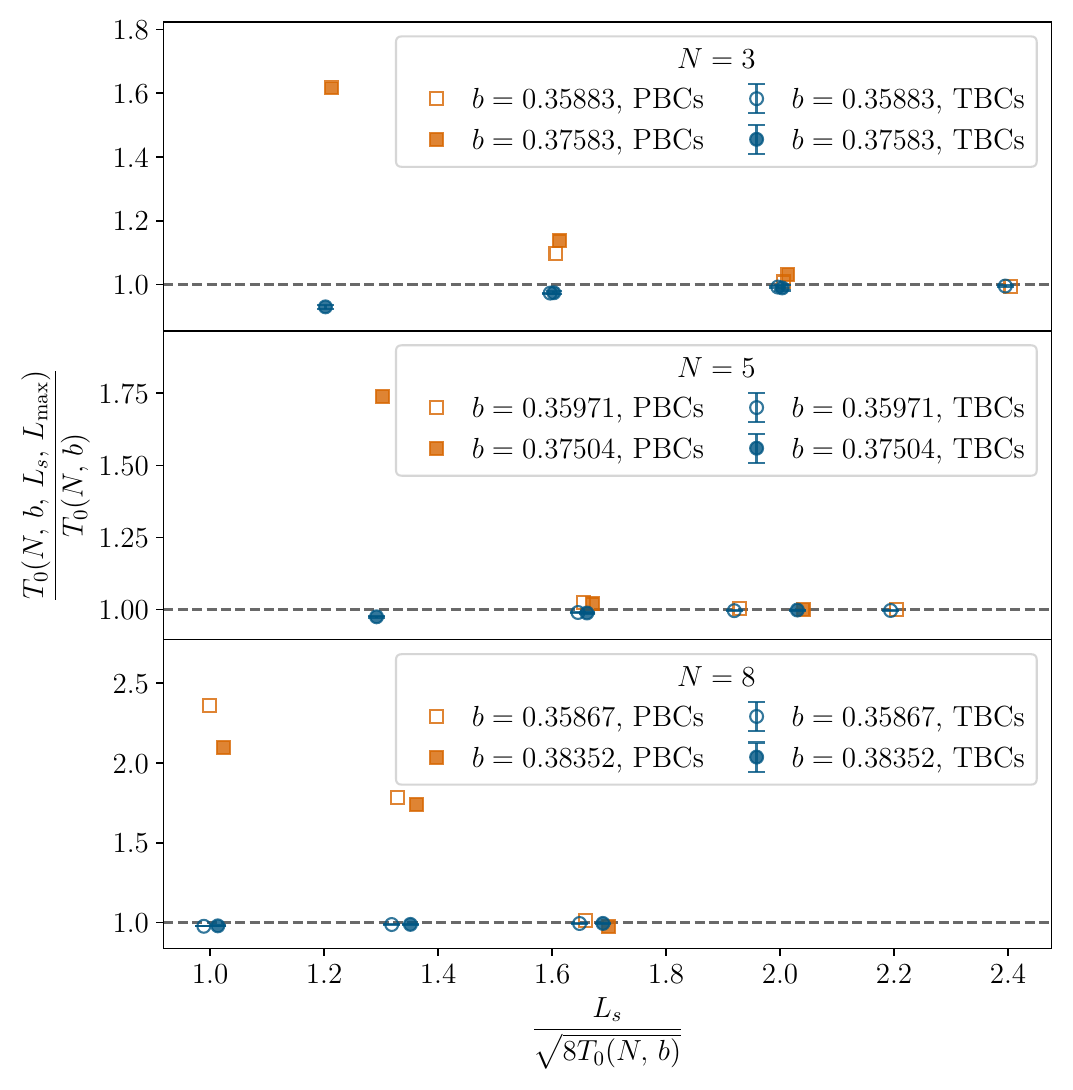}
\caption{Comparison of the dependence of the non-projected scale \tzero on the size $l_s$ of the short plane with twisted (\TBC) and periodic (\PBC) boundary conditions. Finite-volume results are normalized with the thermodynamic-limit extrapolation of $t_0(N,b)$ obtained for \TBC. Results for \PBC are reported without errors because those at small $l_s$ are obtained from a linear extrapolation of \flow beyond the largest time up to which we integrated the flow equation.}
\label{fig:t0+_NbyN_twst_std_all_vs_twist}
\end{figure}

Let us now compare our determinations with previous results. In \cref{t0+_NbyN_twst_std_comb_vs_beta}, we show that our results for the \SU{3} scale \tzero without improvement, reported in \cref{tab:scale-setting-comb-therm-NbyN-twst-std}, are indeed compatible within less than one standard deviations with an interpolation of the results of Ref.~\cite{Giusti:2018cmp} in all cases. For \SU{5}, we can only compare one lattice spacing with Ref.~\cite{Ce:2016awn}, as we explore much finer lattice spacings compared to that study (our coarser lattice spacing is approximately the same as their finest one). Also in this case, our result $t_0/a^2 = 6.643(11)[4]$ is compatible with the interpolation of Ref.~\cite{Ce:2016awn}'s data $t_0/a^2 = 6.6420(16)$ at $\beta=17.985$. For \SU{8}, there is instead no result in the literature to compare with. Overall, \cref{tab:scale-setting-comb-therm-NbyN-twst-std} spells out clearly how our setup allowed us to explore unprecedented regimes, which are out of reach for standard algorithms.

Finally, we conclude our discussion by showing, in \cref{fig:t0+_NbyN_twst_std_all_vs_twist}, the reduction of finite-volume effects achieved thanks to \TBC over \PBC for the scale \tzero. With \TBC and for a short size $l_s \simeq \sqrt{8 t_0} \sim 0.475$ fm, \tzero exhibits a $\sim 10\%$ deviation from the asymptotic value for \SU{3}, which is reduced to $\sim 2\%$ for \SU{8}. In the case of \PBC instead, deviations by $50-100\%$ are obtained for all $N$ at $l_s\simeq\sqrt{8 t_0}$. Asymptotically, boundary conditions do not matter anymore, and the two determinations agree.

\subsection{Scales from a global analysis of the \ensuremath{N}-dependence}\label{sec:results:volume-glob}

\begin{figure}[!t]
\begin{center}
\includegraphics[width=\columnwidth]{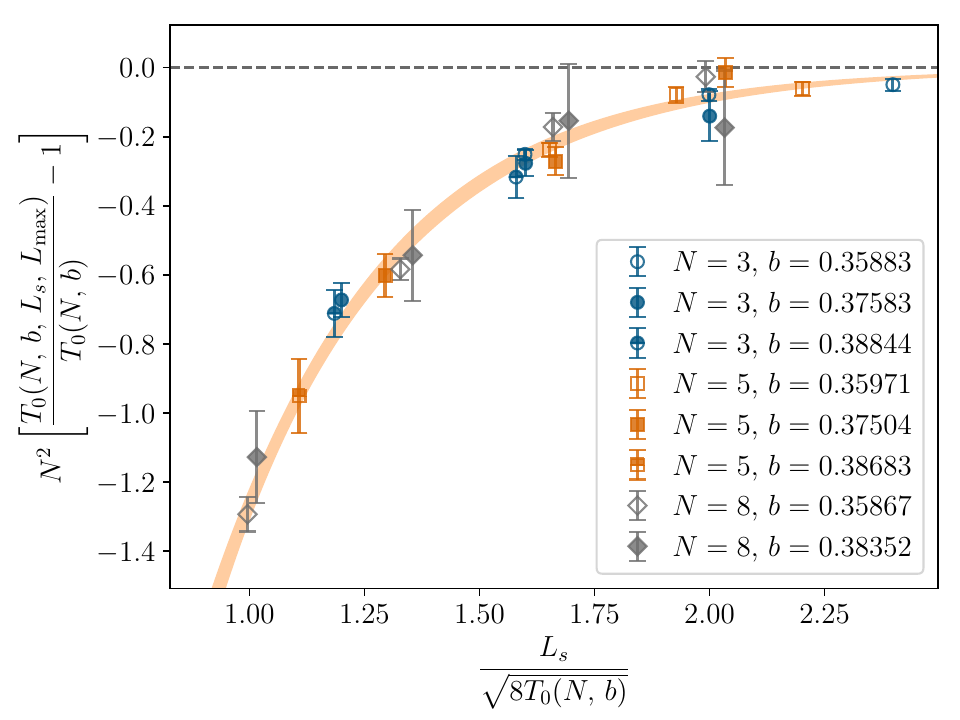}
\caption{Dependence of the non-projected scale \tzero on the size $L_s$ of the short, twisted plane, as in \cref{fig:t0+_NbyN_twst_imp2_anyQ_vs_Ls}, rescaling the finite-volume correction with $N^2$. Results are obtained from the global fit imposing $\hat{A}(N) = A_0/N^2$ in \cref{eq:fit-volume-NbyN-anyQ}.}
\label{fig:t0+_glob_twst_imp2_anyQ_vs_Ls_collapse}
\end{center}
\end{figure}

The results presented in the preceding section lend support to the hypothesis that finite-size effects in the short twisted directions are suppressed by powers of $1/N^2$. This section will undertake a more thorough examination of finite-$N$ and finite-size effects via a global fit, encompassing all the values of $N$. We will proceed as before, by analyzing separately the scales obtained by taking into consideration all topological sectors and the ones derived after projection onto the trivial topology sector. 

In the case of the non-projected scales, we consider again the fit function in \cref{eq:fit-volume-NbyN-anyQ}, but we now enforce the following $N$-dependent form of the prefactor:
\be\label{eq:A_corr}
\hat{A}(N) = \frac{A_0}{N^2+A_1} = \frac{A_0}{N^2} \left[1 - \frac{A_1}{N^2} +\mathcal{O}\left(\frac{1}{N^4}\right)\right] \, .
\ee

\begin{figure}[!t]
\centering
\includegraphics[width=0.9\columnwidth]{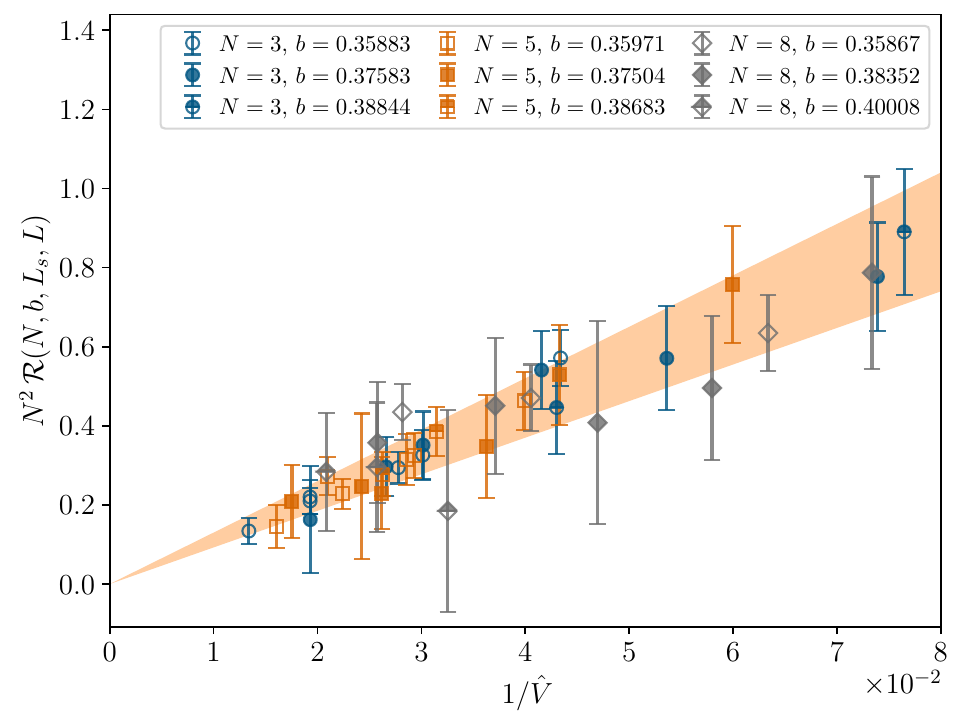}
\includegraphics[width=0.9\columnwidth]{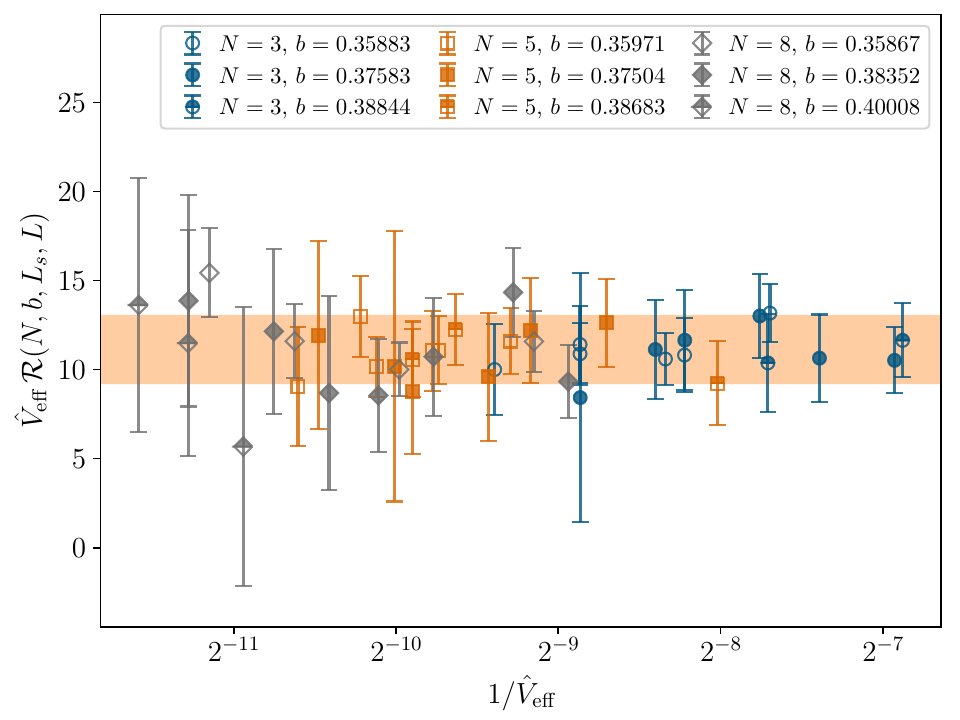}
\caption{Leading finite-volume correction $\mathcal{R}$, defined in \cref{eq:residual-fit-correction-NbyN}, affecting the projected scale \tzeroproj after the subtraction of the subleading exponential term, obtained from the global fit enforcing \cref{eq:A_corr} and \cref{eq:C_corr}. In the top panel, $\mathcal{R}$ is rescaled by $N^2$ and showed as a function of the volume in physical units. In the bottom panel, $\mathcal{R}$ is rescaled by the effective volume to better show the quality of the global fit across all the range of volumes and values of $N$, with the horizontal band representing the fitted value of $C$ in \cref{eq:C_corr}.}
\label{fig:t0+_glob_twst_imp2_Q=0_vs_V_collapse}
\end{figure}

The quality of the global fits is very good, with reduced $\chi^2$ between 0.6 and 1 over 12 degrees of freedom for all the scale definitions. To further check the assumption concerning the $N$-dependence of the finite-size effects, we also performed the global fits for \tzero setting $A_1 = 0$, that is, neglecting subleading corrections to $\hat{A}(N)$, and found a negligible effect on the infinite-volume scales and a still acceptable quality of the fits. To illustrate the goodness of the $1/N^2$ scaling of finite-size effects even with this further constraint, in \cref{fig:t0+_glob_twst_imp2_anyQ_vs_Ls_collapse} we show the quantity
\benn
N^2\left[\frac{T_0(N,b, L_s,L)}{T_0(N,b)} - 1\right]
\eenn
as a function of $\hat{L}_s$ for the improved \tzero. The points, already presented in \cref{fig:t0+_NbyN_twst_imp2_anyQ_vs_Ls} with $N$-by-$N$ fits, are now all fitted together imposing $\hat{A}(N) = A_0 / N^2$. As expected, when appropriately scaled with $N$, all the points collapse on the same curve.

An analogous analysis can be performed for the scales obtained after topological projection. The global fit function is \cref{eq:fit-volume-NbyN-Q=0}, imposing the form in \cref{eq:A_corr} for $\hat{A}(N)$ and similarly:
\be\label{eq:C_corr}
\hat{C}(N) = \frac{C}{N^2} \, ,
\ee
without subleading terms. Again, we obtain very good fit qualities. For the improved \tzeroproj, we show in \cref{fig:t0+_glob_twst_imp2_Q=0_vs_V_collapse} the collapse of all the $N$-dependent curves presented in \cref{fig:t0+_NbyN_twst_imp2_Q=0_vs_V} when the residual volume correction \cref{eq:residual-fit-correction-NbyN} is multiplied by $N^2$.

\begin{figure}[!t]
\centering
\includegraphics[width=\columnwidth]{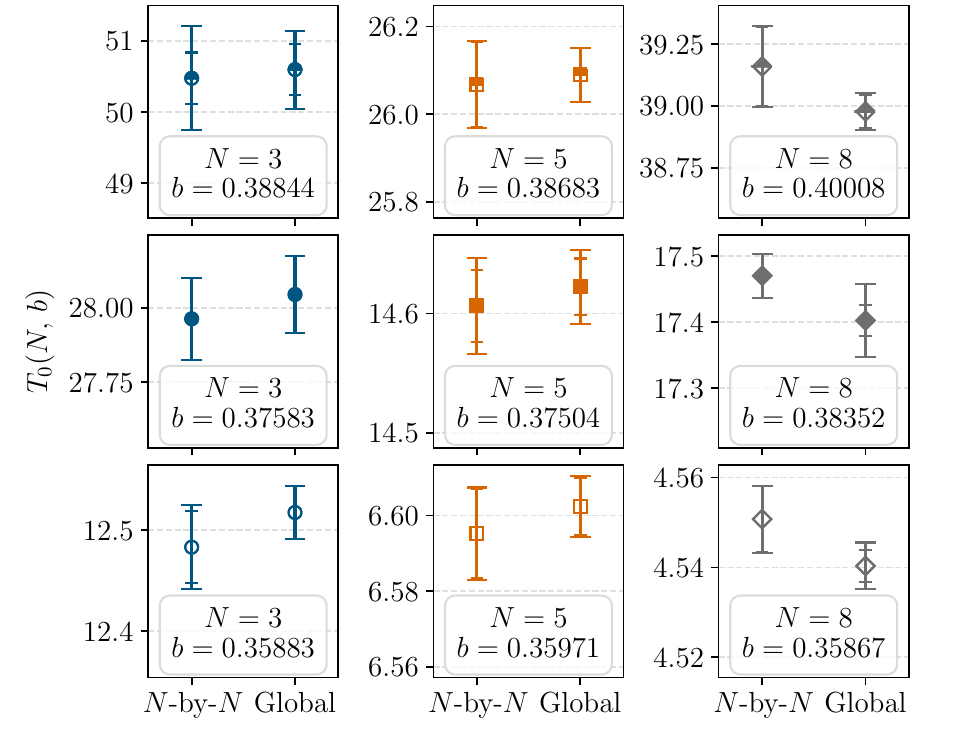}
\caption{Final comparison among results obtained from the $N$-by-$N$ and the global fit strategies to extrapolate towards the thermodynamic limit. Plot refers to improved scales.}
\label{fig:t0+_twst_imp2_final_comparison}
\end{figure}

These results are a confirmation that the expected volume reduction tied to the use of \TBC is attained, and that no residual dependence of finite-size effects in the short twisted directions remains in the large-$N$ limit. The infinite-volume extrapolated scales extracted from these global fits are presented in \cref{tab:scale-setting-comb-therm-glob-twst-imp2} and \cref{tab:scale-setting-comb-therm-glob-twst-std}, and compared with the results obtained in the previous section in \cref{fig:t0+_twst_imp2_final_comparison}. The two determinations agree within errors in every case.

\begin{table}[!t]
\begin{center}
\begin{tabular}{ccccc}
\toprule
\multicolumn{5}{c}{Improved --- Global fit strategy} \\
\midrule
$N$ & $b$ & $t_0/a^2$  & $t_1/a^2$  & $w_0^2/a^2$ \\
\midrule
\multirow{3}{*}[-0.0em]{3} &    0.35883  &     12.518(27)[2]  &     5.2446(55)[15]  &     13.062(56)[9] \\
                           &    0.37583  &      28.04(13)[1]  &      11.745(32)[2]  &      29.24(23)[1] \\
                           &    0.38844  &     50.60(35)[42]  &       21.13(10)[8]  &     52.86(62)[87] \\
\addlinespace[0.5em]
\multirow{3}{*}[-0.0em]{5} &    0.35971  &    6.6024(75)[31]  &     2.6802(13)[10]  &      7.119(19)[5] \\
                           &    0.37504  &    14.622(24)[20]  &     5.9409(48)[59]  &    15.803(54)[29] \\
                           &    0.38683  &     26.089(61)[1]  &      10.599(14)[1]  &      28.20(13)[4] \\
\addlinespace[0.5em]
\multirow{3}{*}[-0.0em]{8} &    0.35867  &    4.5403(36)[37]  &    1.81831(73)[82]  &    4.9612(91)[61] \\
                           &    0.38352  &    17.402(24)[50]  &       6.994(6)[12]  &    19.047(52)[86] \\
                           &    0.40008  &    38.977(66)[36]  &      15.698(16)[7]  &      42.58(14)[8] \\
\bottomrule
\end{tabular}
\caption{Infinite-volume scale-setting results of the global analysis of all values of $N$ with \TBC, flow normalized as in \cref{eq:phi} and improvement of lattice artifacts in \cref{eq:improvement-artifacts}. Results with and without projection onto the \qzero topological sector are combined into a single value with a systematic error (in square brackets), as explained in the text.}
\label{tab:scale-setting-comb-therm-glob-twst-imp2}
\end{center}
\end{table}

\begin{table}[!t]
\begin{center}
\begin{tabular}{ccccc}
\toprule
\multicolumn{5}{c}{Unimproved --- Global fit strategy} \\
\midrule
$N$ & $b$ & $t_0/a^2$  & $t_1/a^2$  & $w_0^2/a^2$ \\
\midrule
\multirow{3}{*}[-0.0em]{3} &    0.35883  &     12.561(26)[3]  &     5.3088(56)[16]  &     13.075(57)[7] \\
                           &    0.37583  &      28.08(13)[1]  &      11.808(32)[2]  &      29.26(23)[1] \\
                           &    0.38844  &     50.61(35)[43]  &       21.19(10)[7]  &     52.91(62)[83] \\
\addlinespace[0.5em]
\multirow{3}{*}[-0.0em]{5} &    0.35971  &    6.6504(74)[30]  &     2.7471(14)[11]  &      7.124(19)[5] \\
                           &    0.37504  &    14.668(24)[19]  &     6.0063(49)[57]  &    15.809(54)[28] \\
                           &    0.38683  &     26.131(61)[1]  &      10.662(14)[1]  &      28.21(13)[4] \\
\addlinespace[0.5em]
\multirow{3}{*}[-0.0em]{8} &    0.35867  &    4.5914(36)[40]  &    1.88740(75)[96]  &    4.9595(92)[60] \\
                           &    0.38352  &    17.451(24)[50]  &       7.059(6)[11]  &    19.048(52)[86] \\
                           &    0.40008  &    39.024(66)[37]  &      15.765(16)[7]  &      42.59(14)[8] \\
\bottomrule
\end{tabular}
\caption{Infinite-volume scale-setting results of the global analysis of all values of $N$ with \TBC, flow normalized as in \cref{eq:phi} and without improvement of lattice artifacts in \cref{eq:improvement-artifacts}. Results with and without projection onto the \qzero topological sector are combined into a single value with a systematic error (in square brackets), as explained in the text.}
\label{tab:scale-setting-comb-therm-glob-twst-std}
\end{center}
\end{table}

\subsection{Discussion of the obtained results}\label{sec:results:continuum-glob}

In this section, we calculate specific ratios of gradient-flow scales and extrapolate them to the continuum and large-$N$ limits. Our motivation is twofold. First, these ratios are of particular utility as conversion factors between scales. Second, some of these large-$N$ ratios can be compared with analogous results recently obtained via twisted volume reduction in Ref.~\cite{Bonanno:2025hzr}. Given the lack of previous finite-$N$ results with which to compare, this is a useful cross-check of our findings.

To demonstrate the effectiveness of the improvement in \cref{eq:improvement-artifacts} in reducing lattice artifacts, we will employ both the improved and unimproved scales. It is important to note that ratios of improved scales are not expected to be free of $\mathcal{O}(a^2)$ lattice artifacts. Besides the cut-off effects associated with the discretization of the energy and canceled at tree-level with \cref{eq:improvement-artifacts}, other corrections arise from the discretization of the simulated action and of the flow equation~\cite{Ramos:2015baa,RamosMartinez:2023tvx}. Thus, we will assume leading ${\cal O}(a^2)$ lattice artifacts for both unimproved and improved ratios.

Ratios of scales are computed with the infinite-volume results of the $N$-by-$N$ analysis discussed in \cref{sec:results:volume-NbyN} and presented in \cref{tab:scale-setting-comb-therm-NbyN-twst-std,tab:scale-setting-comb-therm-NbyN-twst-imp2}. The statistical error assigned to each ratio is estimated by adding in quadrature the statistical errors on the numerator and denominator as if they were uncorrelated. This is a conservative option, which overestimates the uncertainty in the ratios. Concerning systematic errors, we took them into account separately via a bootstrap analysis: each scale is resampled adding a zero-mean bias with variance equal to the systematic error. Then, the systematic error on the extrapolation is computed via the bootstrap mean of \cref{eq:syst}, considering the difference between the extrapolation obtained without systematic errors and the ones obtained in each bootstrap sample. We found that the resulting systematics are completely negligible with respect to statistical errors, which dominate the total errors on our final results.

\begin{table}[!t]
\centering
\begin{tabular}{ccccc}
\toprule
\multicolumn{5}{c}{$N$-by-$N$ fit strategy} \\
\midrule
 & \multicolumn{2}{c}{Unimproved} & \multicolumn{2}{c}{Improved} \\
\cmidrule(lr){2-3}\cmidrule(lr){4-5}
$N$ & $t_1/t_0$ & $w_0^2/t_0$ & $t_1/t_0$ & $w_0^2/t_0$ \\
\midrule
3  &  0.4188(33)  &   1.044(14)  &  0.4187(33)  &   1.043(14) \\
5  &  0.4067(15)  &  1.0836(94)  &  0.4067(15)  &  1.0827(94) \\
8  &  0.4013(10)  &  1.0976(66)  &  0.4015(10)  &  1.0973(65) \\
$\infty$  &  0.3985(13)  &  1.1063(76)  &  0.3988(13)  &  1.1060(75) \\
\bottomrule
\end{tabular}
\caption{Continuum scale ratios from the $N$-by-$N$ analysis of unimproved and improved scales.}
\label{tab:scale-setting-cont-NbyN-twst}
\end{table}

\begin{figure}[!t]
\centering
\includegraphics[width=\columnwidth]{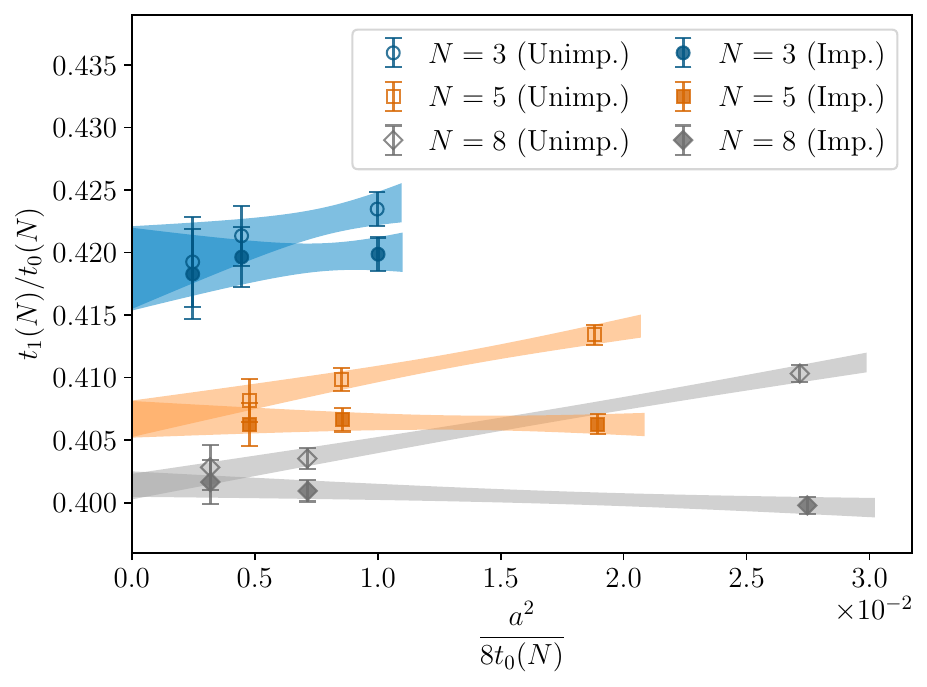}
\caption{Continuum extrapolations of the ratio $t_1/t_0$ for several values of $N$, with and without the improvement in \cref{eq:improvement-artifacts}. Each dataset at different value of $N$, as well as improved/unimproved datasets, have been fitted separately.}
\label{fig:ratio_t1+t0+_vs_a_NbyN}
\end{figure}

Similarly to the analysis of finite-volume effects, we adopt two fitting strategies to determine the continuum and large-$N$ limits. In one case, we first perform the continuum extrapolation of each ratio for each value of $N$ separately, then we extrapolate the continuum results to $N = \infty$. The fit function for the continuum extrapolations is
\be\label{eq:fit-artifacts-NbyN}
\frac{S_1(N,b)}{S_2(N,b)} = R(N)\left[1 + \frac{D(N)}{8S_2(N,b)}\right] \, ,
\ee
where $1/S_2(N,b)$ measures $a^2$ in units of the denominator scale. The fit parameters are $D(N)$ and $R(N)$, the ratio between the scales $S_1$ and $S_2$ in the continuum limit. The large-$N$ limit $R$ of the ratios is then obtained by fitting
\be\label{eq:fit-largeN-NbyN}
R(N) = R\left(1 + \frac{C}{N^2}\right) \, ,
\ee
that is, assuming leading $1/N^2$ corrections. In \cref{fig:ratio_t1+t0+_vs_a_NbyN} we show the continuum extrapolations of the ratio $t_1/t_0$ for all values of $N$, with and without improvement of lattice artifacts. As it is evident, the improved scale determinations demonstrate a reduced lattice-spacing dependence compared to the unimproved ones. In particular, the prefactor $D(N)$ turns out to be very small and practically compatible with zero, possibly due to an accidental cancellation of the $\mathcal{O}(a^2)$ corrections on the scales in the ratio, at least up to our precision. The difference between the two continuum extrapolations is negligible with respect to the statistical error. These observations hold for all values of $N$ and also for the other ratios of scales we computed. The continuum and large-$N$ limits for the ratios of \tone and \wzero over \tzero are presented in \cref{tab:scale-setting-cont-NbyN-twst}. Other ratios, best-fit parameters and the quality of the fits are reported in the Supplemental Material~\cite{suppmat}.

\begin{figure}[!t]
\centering
\includegraphics[width=0.95\columnwidth]{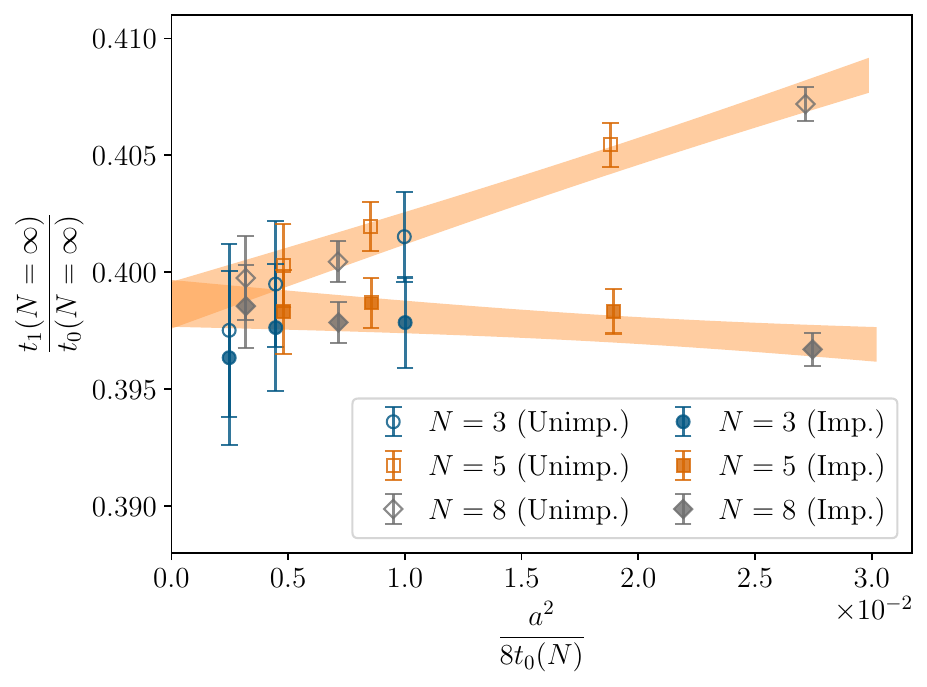}
\caption{Continuum extrapolation of the large-$N$ ratio $t_1/t_0$, with and without the improvement in \cref{eq:improvement-artifacts}. All values of $N$ are fitted together in a global fit of both finite-$N$ effects and lattice artifacts according to the fit function in \cref{eq:fit-artifacts-glob}.}
\label{fig:ratio_t1+t0+_vs_a_collapse_glob}
\end{figure}

\begin{figure}[!t]
\centering
\includegraphics[width=0.95\columnwidth]{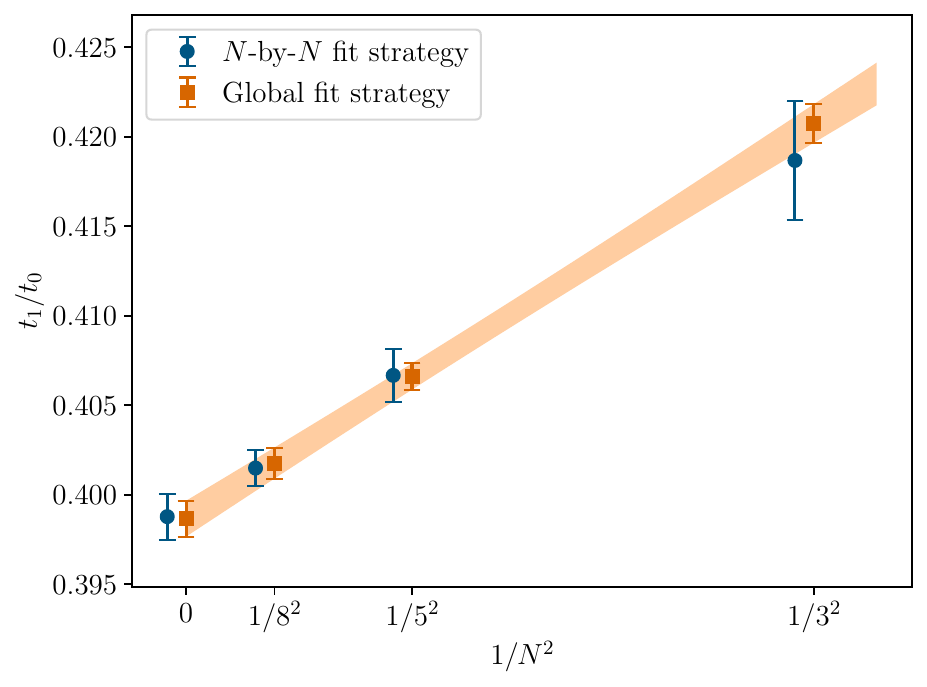}
\caption{Large-$N$ extrapolation of the continuum-extrapolated ratio $t_1/t_0$ with the improvement in \cref{eq:improvement-artifacts}. All values of $N$ are fitted together in a global fit of both finite-$N$ effects and lattice artifacts. The points resulting from the $N$-by-$N$ fits presented in \cref{fig:ratio_t1+t0+_vs_a_NbyN} and their large-$N$ extrapolation are also shown.}
\label{fig:ratio_t1+t0+_vs_N}
\end{figure}

The results of the $N$-by-$N$ fits show a negligible dependence on $N$ of the prefactor controlling the size of lattice artifacts, $D(N)$. This observation motivates our second strategy, in which we perform a global fit of finite-$a$ and finite-$N$ effects according to the following function:
\be\label{eq:fit-artifacts-glob}
\frac{S_1(N,b)}{S_2(N,b)} = R\left(1 + \frac{C}{N^2}\right)\left[1 + \frac{D}{8S_2(N,b)}\right] \, .
\ee
As an example of the very good quality of the global fits, we show in \cref{fig:ratio_t1+t0+_vs_a_collapse_glob} the quantity
\benn
\frac{S_1(N,b)}{S_2(N,b)} \left(1 + \frac{C}{N^2}\right)^{-1} \, ,
\eenn
representing the lattice-spacing-dependence in the large-$N$ limit, for the ratio of \tone and \tzero. The points at different values of $N$, already presented in \cref{fig:ratio_t1+t0+_vs_a_NbyN}, collapse into two single curves, corresponding to the ratios of scales with and without improvement. The two strategies give perfectly compatible results at each finite $N$ and in the large-$N$ limit, as illustrated in \cref{fig:ratio_t1+t0+_vs_N}, where we show the $N$-dependence of the improved and continuum-extrapolated ratio. Results from the global fit are reported in \cref{tab:scale-setting-cont-glob-twst}.

\begin{table}[!t]
\centering
\begin{tabular}{ccccc}
\toprule
\multicolumn{5}{c}{Global fit strategy} \\
\midrule
 & \multicolumn{2}{c}{Unimproved} & \multicolumn{2}{c}{Improved} \\
\cmidrule(lr){2-3}\cmidrule(lr){4-5}
$N$ & $t_1/t_0$ & $w_0^2/t_0$ & $t_1/t_0$ & $w_0^2/t_0$ \\
\midrule
3  &   0.4204(11)  &  1.0425(54)  &   0.4207(11)  &  1.0411(54) \\
5  &  0.40643(74)  &  1.0833(47)  &  0.40661(74)  &  1.0825(46) \\
8  &  0.40165(87)  &  1.0973(56)  &  0.40177(87)  &  1.0967(55) \\
$\infty$  &   0.3986(10)  &  1.1062(64)  &   0.3987(10)  &  1.1058(63) \\
\bottomrule
\end{tabular}
\caption{Continuum scale ratios from the global analysis of unimproved and improved scales.}
\label{tab:scale-setting-cont-glob-twst}
\end{table}

\begin{figure}[!t]
\centering
\includegraphics[width=\columnwidth]{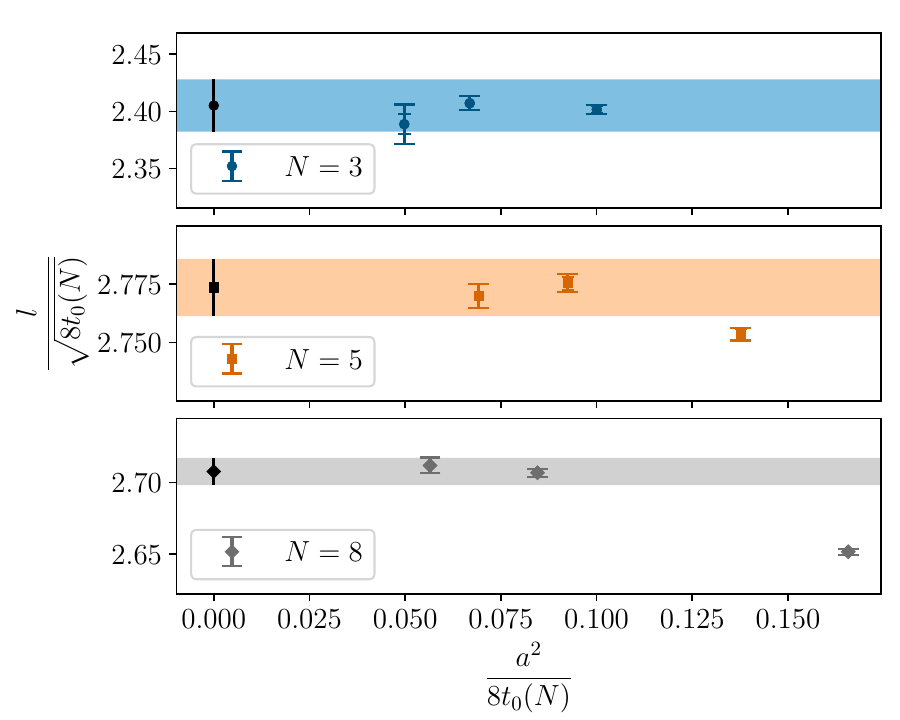}
\caption{Lattice sizes in units of the improved \tzero for the simulations in \cref{tab:tuned_simulations}, with parameters calibrated via the TGF coupling. Colored bands represent a fit to a constant excluding the coarsest lattice spacing with errors multiplied by 4.}
\label{fig:tuned_lcps}
\end{figure}

Our $N=3$ determination of $t_0/w_0^2=0.9588(51)$ agrees perfectly with the recent result $t_0/w_0^2=0.9579(35)$ of~\cite{Holland:2025fsa} obtained from simulations with a machine-learned classically perfect action. As a further consistency check of our calculations, we also computed the $N=\infty$ ratios $t_0/t_0^\prime$ and $t_1^\prime/t_0^\prime$. These can be compared with the same quantities recently obtained within the Twisted Eguchi--Kawai (TEK) model~\cite{Bonanno:2025hzr} in the large-$N$ limit. We find $t_0/t_0^\prime = 1.1432(42)$ and $t_1^\prime/t_0^\prime=0.3744(11)$, in good agreement with TEK results $t_0/t_0^\prime = 1.140(3)$ and $t_1^\prime/t_0^\prime=0.3953(125)$.

Finally, let us conclude our discussion by using our scale-setting results to compute the values of the lattice size in physical units for the simulation parameters calibrated via the TGF coupling reported in \cref{tab:tuned_simulations}. For each $N$, those simulation points should all correspond to the same value of $l=aL/\sqrt{8t_0}$, modulo small lattice artifacts. We show results for $l$ in \cref{fig:tuned_lcps}. As it can be observed, our results for the scale setting support our theoretical expectations. In particular, using $\sqrt{8t_0} \simeq 0.475$ fm~\cite{Giusti:2018cmp}, we find $l \simeq 1.14$ fm for $N=3$ and $l \simeq 1.30$ fm for $N=5,8$, as expected.

\section{Conclusions}\label{sec:conclusions}

In this paper we have presented scale-setting results for large-$N$ Yang--Mills theory obtained via the gradient flow. This is the first step within a larger project whose ultimate goal is to achieve the first lattice determination of the large-$N$ Yang--Mills $\Lambda$-parameter using the step scaling method.

Scale setting at large-$N$ for step scaling applications requires to explore a regime of lattice Yang--Mills theories (large values of $N$, fine lattice spacings), where topological freezing is extremely severe. In this study we tackled this problem by adopting the PTBC to effectively reduce autocorrelation times of topological quantities, and Twisted Boundary Conditions (\TBC) to exploit large-$N$ volume reduction and keep finite-size effects under control. 

Our setup allowed us to reliably compute gradient-flow scales in regimes which are completely out of reach for standard algorithms. For $N=3$, we nontrivially verified consistency of our ergodic determinations with previous Master Field simulation results. In all cases, we were capable of achieving control over systematic finite-size effects, both of standard nature and due to the fixed topological background that would be obtained in the presence of topological freezing. In particular, we were able to clearly probe the exponentially-suppressed finite volume effects dominating scale setting quantities in the presence of a properly-sampled topology, and the powerlike ones dominating topology-projected scale determinations. Finally, we provided clear numerical evidence of the expected $1/N^2$ suppression of finite-volume effects in the short, twisted lattice size $l_s$ expected on general theoretical grounds by virtue of large-$N$ volume independence.

We concluded our study providing further cross checks of our results. For $N=3$, our ratio of gradient flow scales $t_0/w_0^2$ perfectly agree with the recent ones achieved in~\cite{Holland:2025fsa} by means of a machine-learned classically perfect action. In the large-$N$ limit, our ratios of gradient-flow scales $t_0/t_0^\prime$ and $t_0/t_1^\prime$ are found in agreement with those determined using the TEK model with $N$ as large as 841. We also verified that lattice parameters calibrated to have the same renormalized strong coupling in the so-called Twisted Gradient Flow scheme indeed corresponded to the same physical value of the lattice size $\hat{l}=aL/\sqrt{8t_0}$ (up to small lattice artifacts).

In the next future, we will employ the results presented in this study to set the scale in our forthcoming step scaling study of the large-$N$ Yang--Mills $\Lambda$-parameter. Indeed, step scaling determines $\Lambda$ in units of an arbitrary energy scale $\muhad$. Our scale-setting results will thus allow the conversion of $\muhad$, and thus of $\Lambda$, into units of $\sqrt{8t_0}$. This investigation will be presented in a forthcoming publication.

\acknowledgments 
This work is partially supported by the Spanish Research Agency (Agencia Estatal de Investigaci\'on) through the IFT Centro de Excelencia Severo Ochoa Grant No. CEX2020-001007-S and, partially, by Grant No. PID2021-127526NB-I00 and PID2024-160152NB-I00, both funded by MCIN/AEI/10.13039/501100011033. This work has also been partially supported by the project ”Non-perturbative aspects of fundamental interactions, in the Standard Model and beyond” funded by MUR, Progetti di Ricerca di Rilevante Interesse Nazionale (PRIN), Bando 2022, Grant No. 2022TJFCYB (CUP I53D23001440006). This work has also been partially supported by the U.S. Department of Energy, Office of Science, Office of Nuclear Physics under Contract No. DE-SC0012704 and by the Scientific Discovery through Advanced Computing (SciDAC) award "Fundamental Nuclear Physics at the Exascale and Beyond" and the Topical Collaboration in Nuclear Theory "Heavy-Flavor Theory (HEFTY) for QCD Matter". Numerical calculations have been performed partially on the \texttt{Leonardo} machine at Cineca, based on the agreement between INFN and Cineca, under the projects INF24\_npqcd and INF25\_npqcd, and partially on the \texttt{Finisterrae~III} cluster at CESGA (Centro de Supercomputaci\'on de Galicia).

\section*{Data availability}

The data that support the findings of this article are openly available at~\cite{suppmat}. Further data available upon reasonable request.

\appendix

\section*{Appendix}

\section{Topology sampling and autocorrelations with the PTBC algorithm}\label{sec:appendix}

\begin{table}[!t]
\centering
\begin{tabular}{ccccc}
\toprule
$N$ & $b$ & $N_r$ & $\tau_0(Q^2)$ & $\tau(Q^2)$ \\
\midrule
\multirow{3}{*}[-0.0em]{3} &    0.35883  &  18  &  2.4(1.0)$\cdot10^2$  &  4.3(1.7)$\cdot10^3$ \\
                           &    0.37583  &  34  &  1.1(0.4)$\cdot10^3$  &  3.6(1.6)$\cdot10^4$ \\
                           &    0.38844  &  54  &  1.3(0.6)$\cdot10^3$  &  7.3(3.1)$\cdot10^4$ \\
\addlinespace[0.5em]
\multirow{3}{*}[-0.0em]{5} &    0.35971  &  21  &  3.2(1.1)$\cdot10^2$  &  6.8(2.3)$\cdot10^3$ \\
                           &    0.37504  &  32  &  6.9(2.9)$\cdot10^2$  &  2.2(1.0)$\cdot10^4$ \\
                           &    0.38683  &  44  &  $\gtrsim$ 3.8(1.0)$\cdot10^2$  &  1.7(0.4)$\cdot10^4$ \\
\addlinespace[0.5em]
\multirow{2}{*}[-0.0em]{8} &    0.35867  &  18  &  7.2(2.4)$\cdot10^2$  &  1.2(0.4)$\cdot10^4$ \\
                           &    0.38352  &  46  &  $\gtrsim$ 7.3(3.9)$\cdot10^2$  &  3.4(1.8)$\cdot10^4$ \\
\bottomrule
\end{tabular}
\caption{Summary of auto-correlation times of the squared topological charge $\tau_0(Q^2)$ of the periodic replica, expressed in units of lattice sweeps. These times refer to the largest simulated volume at each $(N,b)$. We also report $\tau(Q^2) = N_r \tau_0(Q^2)$, a useful figure of merit that also keeps into account the replica overhead. In some cases, only lower bound could be obtained.}
\label{tab:tauQ}
\end{table}

The present study was possible thanks to the improved scaling of the autocorrelation time of the topological charge achieved via the PTBC algorithm, studied in detail in Ref.~\cite{Bonanno:2025eeb}. These are reported in Tab.~\ref{tab:tauQ}. In order to check that $Q$ has been sampled correctly, let us compute $\langle{Q^2}\rangle$ for a representative ensemble (a detailed study of Yang--Mills topology with TGF and PTBC will be left for a future publication). Let us select the finest lattice spacing of SU(5). This simulation has been performed at a value $\ell/\sqrt{8t_0}\simeq 2.97$ of the long lattice size with negligible finite-$\ell$ effects for all the other simulation points, as shown in \cref{fig:t0+_NbyN_twst_imp_all_vs_L}, and a value $\ell_s/\sqrt{8t_0}\simeq 1.12$ of the short lattice size for which the finite-volume correction in \cref{eq:fit-volume-NbyN-anyQ} is $t_0(\ell_s)/t_0 = 0.957(7)$, as shown in \cref{fig:t0+_NbyN_twst_imp2_anyQ_vs_Ls}. This volume is small enough to run our single-node PTBC implementation using the resources available to us. The finite-volume result for the improved $t_0$ scale is $t_0(\ell_s)/a^2 = 25.08(11)$. Thus, the infinite-volume prediction is $t_0/a^2 = 26.20(23)$, in agreement with the determination $26.10(11)$ obtained from the infinite-volume extrapolation of only the frozen simulations.

\begin{figure}[!t]
\centering
\includegraphics[width=\columnwidth]{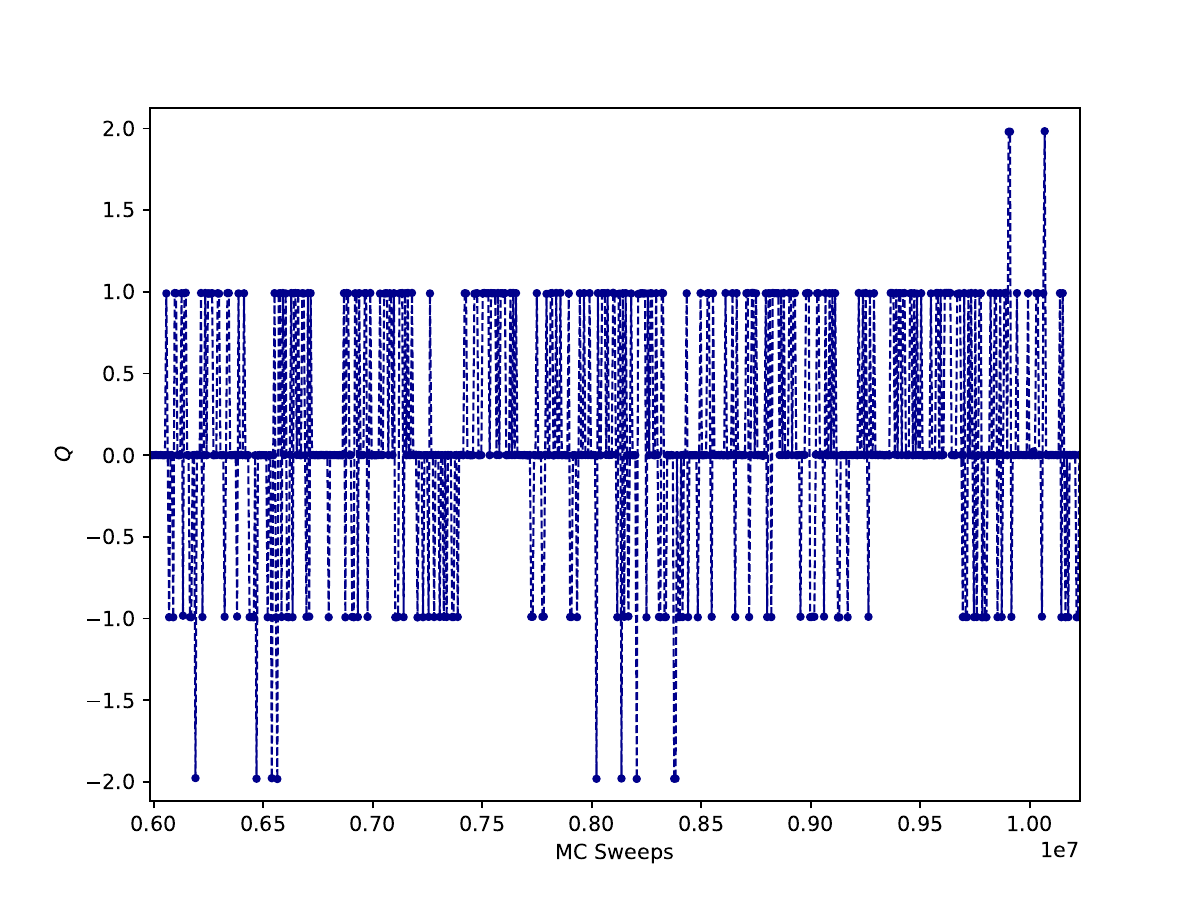}
\caption{Fraction of the Monte Carlo history for $N=5$, $b=0.38683$ (finest lattice spacing), $L_s/\sqrt{8t_0}\simeq 1.12$, $L/\sqrt{8t_0}\simeq 2.97$. The shown fraction corresponds to one-third of the total statistics.}
\label{fig:charge_history}
\end{figure}

In \cref{fig:charge_history} we show part of the Monte Carlo history of the topological charge attained with the PTBC algorithm. Using the continuum and infinite-volume value of $t_0^2\chi$ for $N=5$ of~\cite{Ce:2016awn} and our value of the volume in units of $t_0$, we estimate $\langle Q^2\rangle=\chi V\simeq 0.48(1)$. Our run yields $\langle Q^2\rangle=0.434(38)$, quite close to the thermodynamic limit despite being outside the asymptotic regime (the correction on $t_0$ is $\simeq 4\%$), proving that the PTBC algorithm allows to sample the topological charge correctly.

\FloatBarrier

\bibliographystyle{apsrev4-2}
\bibliography{biblio}

\end{document}